\documentclass[journal=jctcce,manuscript=article]{achemso}
\setkeys{acs}{articletitle=true}

\usepackage[version=3]{mhchem} 
\usepackage{mciteplus}
\usepackage{rotating}
\usepackage{amscd}
\usepackage{color}
\usepackage{subcaption}
\usepackage[usenames,dvipsnames]{xcolor}
\usepackage{setspace}
\usepackage{xspace}

\newcommand*{\kcal}{kcal\,mol$^{-1}$\xspace}

\newcommand*{\degrees}{$^{\circ}$\xspace}

\usepackage{xspace}

\usepackage{hyperref}
\hypersetup{
    colorlinks,
    linkcolor={black},
    citecolor={black},
    urlcolor={blue}
}
\usepackage{cleveref}
\makeatletter
\newcommand*{\addFileDependency}[1]{
  \typeout{(#1)}
  \@addtofilelist{#1}
  \IfFileExists{#1}{}{\typeout{No file #1.}}
}
\makeatother

\hypersetup{colorlinks=false,linkcolor=black}

\author{Johannes Karwounopoulos$^{\#}$}
\affiliation{Exscientia, Schr\"odinger Building, Oxford Science Park, Oxford, UK}
\author{Mateusz Bieniek$^{\#}$}
\affiliation{Exscientia, Schr\"odinger Building, Oxford Science Park, Oxford, UK} 
\author{Zhiyi Wu}
\affiliation{Exscientia, Schr\"odinger Building, Oxford Science Park, Oxford, UK}
\author{Adam L. Baskerville}
\affiliation{Exscientia, Schr\"odinger Building, Oxford Science Park, Oxford, UK}
\author{Gerhard K\"onig}
\affiliation{Exscientia, Schr\"odinger Building, Oxford Science Park, Oxford, UK}
\email{gkoenig@exscientia.co.uk}
\author{Benjamin P. Cossins}
\affiliation{Exscientia, Schr\"odinger Building, Oxford Science Park, Oxford, UK}
\author{Geoffrey P. F. Wood}
\affiliation{Exscientia, Schr\"odinger Building, Oxford Science Park, Oxford, UK}

\title[]{An evaluation of machine learning/molecular mechanics end-state corrections with mechanical embedding to calculate relative protein--ligand binding free energies}

\begin{document}
\maketitle
\def\thefootnote{\#}\footnotetext{Contributed equally}


\begin{abstract}
The development of machine-learning (ML) potentials offers significant accuracy improvements compared to molecular mechanics (MM) because of the inclusion of quantum-mechanical effects in molecular interactions. However, ML simulations are several times more computationally demanding than MM simulations, so there is a trade-off between speed and accuracy.  One possible compromise are hybrid machine learning/molecular mechanics (ML/MM) approaches with mechanical embedding that treat the intramolecular interactions of the ligand at the ML level and the protein-ligand interactions at the MM level. Recent studies have reported improved protein-ligand binding free energy results based on ML/MM with mechanical embedding, arguing that intramolecular interactions like torsion potentials of the ligand are often the limiting factor for accuracy. This claim is evaluated based on 108 relative binding free energy calculations for four different benchmark systems. As an alternative strategy, we also tested a tool that fits the MM dihedral potentials to the ML level of theory. Overall, the relative binding free energy results from MM with Open Force Field 2.2.0, MM with ML-fitted torsion potentials, and the corresponding ML/MM end-state corrected simulations show no statistically significant differences in the mean absolute errors (between 0.8 and 0.9\,\kcal{}). This can probably be explained by the usage of the same MM parameters to calculate the protein-ligand interactions. Therefore, a well-parameterized force field is on a par with simple mechanical embedding ML/MM simulations for protein-ligand binding. In terms of computational costs, the reparametrization of poor torsional potentials is preferable over employing computationally intensive ML/MM simulations of protein-ligand complexes with mechanical embedding. Also, the refitting strategy leads to lower variances of the protein-ligand binding free energy results than the ML/MM end-state corrections. For free energy corrections with ML/MM, the results indicate that better convergence and more advanced ML/MM schemes will be required for applications in computer-guided drug discovery.   
  
\end{abstract}

\section{Introduction}
\label{sec:introduction}

Relative binding free energy (RBFE) calculations have become a standard tool in computational drug discovery,\cite{williams2018free,homeyer2014binding,cournia2017relative,muegge2023recent, Karwounopoulos2022}  where rigorous physics-based predictions of protein-ligand binding serve both to enrich the number of active compounds for experimental testing,\cite{shirts2010free,mobley2012perspective,koenig2021rational,xue2023development} and to provide machine-learning drug-discovery pipelines with additional data.\cite{burger2024fep}   
Nowadays, RBFE calculations typically yield root mean square errors between $0.4$ and $4.3$\,\kcal{} relative to experimental protein-ligand binding affinities,\cite{hahn2024current,lee2020alchemical,schindler2020large} depending on the quality of the underlying protein structure, sampling, and force field.\cite{van2018validation}

A major challenge for RBFE calculations in computational drug discovery is the accurate  description of molecular interactions from the vast space of drug-like chemical compounds. Uncommon chemical groups  might not be well-supported by force fields, which can lead to large errors. Two possible strategies to address this problem are a) the reparametrization of the force field based on quantum-mechanical (QM) calculations, or b) using a hybrid quantum mechanics/molecular mechanics (QM/MM) approach to describe the ligand.\cite{Gao1992,Gao1993,Luzhkov1992a,Wesolowski1994,Gao1997} One of the most simple QM/MM techniques is mechanical embedding, where the intramolecular interactions of the ligand are treated at the QM level, while the environment and the ligand-environment interactions are calculated at the MM level. Unfortunately, it is rarely possible to perform free energy calculations directly at the QM level of theory,\cite{Reddy2011, Cui2009Multi-ScaleSCC-DFTB,qmnbb} because of the high computational demands and the need to implement dummy atoms and soft core potentials for alchemical transformations. To address these challenges, indirect free energy methods that employ end-state corrections have been developed.\cite{Gao1992,Gao1993,Luzhkov1992a,Wesolowski1994,Gao1997,Valiev2008CombinedSolution,Zheng2008a,beierlein11,heimdal12,fox13,qmnbb,qmmmsampl4,sampson15,cave-ayland15,koenig16a,qmsampl5imp,Giese2019DevelopmentMethods,qmmmsampl5,Cui2021}
Instead of performing the entire alchemical transformation at the QM level, the free energy difference is first calculated at the MM level. Then, the free energy differences between the MM and the QM energy surfaces are computed and added as correction terms to the MM free energy difference. For the RBFE between two ligands A and B at the QM level of theory ($\Delta \Delta G^{\mathrm{QM}}_{\mathrm{A}\rightarrow\mathrm{B}}$), this corresponds to:
\begin{equation}
\Delta \Delta G^{\mathrm{QM}}_{\mathrm{A}\rightarrow\mathrm{B}}= \Delta \Delta G^{\mathrm{MM}}_{\mathrm{A}\rightarrow\mathrm{B}} -  \Delta G_{\mathrm{A}}^{\mathrm{MM}\rightarrow\mathrm{QM}} + \Delta G_{\mathrm{B}}^{\mathrm{MM}\rightarrow\mathrm{QM}}
\end{equation}
Here, $\Delta \Delta G^{\mathrm{MM}}_{\mathrm{A}\rightarrow\mathrm{B}}$ is the RBFE between two ligands at the MM level of theory, while $\Delta G_{\mathrm{A}}^{\mathrm{MM}\rightarrow\mathrm{QM}}$ and $\Delta G_{\mathrm{B}}^{\mathrm{MM}\rightarrow\mathrm{QM}}$ are the end-state corrections from MM to QM for each ligand. In several instances, free energy calculations with QM has substantially  
improved the accuracy of the results under blind conditions.\cite{qmmmsampl4,qmmmsampl5,qmsampl5imp,pickard16,koenig2021rational} 

The main challenge of end-state corrections is the convergence of free energy calculations between the MM and QM.\cite{multiconv} If the MM energy surface is not representative of the QM energy surface, most of the MM sampling is conducted in regions of phase space that have a very low probability in QM, which lowers the number of effective samples. In many cases, the stiffest degrees of freedom in MM, the bond lengths and angles, slightly differ from the QM level of choice.\cite{qmdrude,giese2019development,konig2020faithfulness} To resolve this convergence issue, reparametrization techniques\cite{hudson2018accelerating,qmdrude,konig2020faithfulness,giese2019development,bespokefit}  and nonequilibrium switching (NEQ) methods\cite{Kearns2017ComputingOpportunities, qmjarzynski}  can be employed.

ML approaches trained on QM data have recently been shown to reliably reproduce QM potential energy surfaces, striking a balance between accuracy and speed. \cite{Kulichenko2021,dral2020quantum} A prominent example for such ML approaches is the ANI family of neural network potentials, with {ANI-2x} being particularly popular for tasks like conformer generation and chemical property prediction.\cite{ani1, ani1ccx, ani2x, Park2022, Han2023} ANI-2x was trained with an active learning strategy on $8.9$~million molecular conformations, using the $\omega$B97X/6-31G* level of theory. 
Recently, the use of ANI-2x has been reported for protein-ligand binding calculations using RBFE simulations\cite{rufa2020towards,sabanes2024enhancing} and the linear interaction energy method.\cite{Akkus2022} Refs.~\citenum{rufa2020towards} and \citenum{sabanes2024enhancing} employed an indirect free energy cycle with end-state corrections,\cite{tkaczyk2024reweighting}  where the ANI-2x potential served as the QM potential to correct free energies calculated at the MM level. A similar end-state correction approach was also recently used to calculate hydration free energies.\cite{karwounopoulos2023insights} Therefore, the employment of ML/MM approaches instead of QM/MM approaches to calculate free energy differences is an attractive choice. 

Another possible strategy to employ ML potentials to improve the reliability of protein-ligand binding predictions for diverse chemical spaces is the reparametrization of existing force fields.\cite{qubekit,hudson2018accelerating, qmdrude, konig2020faithfulness, giese2019development} This approach has also been termed bespoke parametrization,\cite{bespokefit,qubekit} or tuned force fields. A recent report indicates that tuned force fields can be on a par with ML/MM simulations in terms of energies and conformational sampling,\cite{morado2023does} which makes the reparametrization approach attractive, as MM simulations are computationally less demanding than ML/MM simulations. Results for drug-like molecules show that the ANI-ML potentials are capable of closely reproducing the  underlying QM torsional potential energy surfaces,\cite{lahey2020benchmarking} which indicates that ML calculations could be employed for the fitting of torsion potentials in an automated computational drug discovery pipeline.     

Here, both ML potentials for the reparametrization of force fields and the use of end-state corrections based on ML/MM simulations with ANI-2x are evaluated for the calculation of RBFE of protein-ligand systems in drug discovery. Four benchmark systems TYK2,\cite{tyk2a,tyk2b} CDK2,\cite{cdk2} JNK1,\cite{jnk1} and P38\cite{p38} from Ref.~\citenum{Wang2015} are employed for to test the accuracy of the methods based on $108$ RBFE results. In the following sections, the methodological details of the RBFE calculations with MM, the torsion potential fitting with the in-house Torpenter tool, and the end-state corrections are described. Following this, the RBFE protocols are verified based on reported results in the literature. The effect of the parameter refitting with ML and end-state corrections with ML/MM  are discussed based on the mean absolute errors of the RBFE results. Finally, a set of recommendations and guidelines are laid out.


\label{sec:methods}
\section{Methods}
\subsection{Relative binding free energy calculations}

RBFE calculations were conducted for $108$ ligand pairs of the benchmark systems TYK2, CDK2, JNK1, and P38.\cite{Wang2015} Four ligands had to be removed because the ANI-2x model is not parameterized for Br (ligand 17 in CDK2, and ligands 17124\_1 and 18636\_1 of JNK1) or charged molecules (ligand 18639\_1 in JNK1). The RBFE calculations employed the standard thermodynamic cycles,\cite{tem84a} which involve free energy simulations of the protein-ligand complex (bound leg) and the ligand in solution (free leg). Ligand mappings were generated with the in-house \texttt{Pertmapper} tool,\cite{pertmapper} and the ligand network was constructed with \texttt{LOMAP}.\cite{lomap} Alchemical transformations were prepared according to the recommendations by Fleck, Wieder, and Boresch,\cite{fleck2021dummy} using an in-house strain removal algorithm. The end states were equilibrated separately using the robust protocol of Roe and Brooks.\cite{roe2020protocol}  The alchemical system was divided into $12$ $\lambda$-windows: the first six windows were started from the equilibration of the initial state ($\lambda=0$) and the final six windows from the structure of the final state ($\lambda=1$). For each $\lambda$-state, the energy was minimized with 100 steps of steepest descent. This was followed by another equilibration phase of $0.04$\,ns with a $1$\,fs timestep at $298$\,K, utilizing the Langevin integrator and a Monte Carlo barostat set to $1$\,atm.\cite{mcbarostat} The production run lasted $2$\,ns with a $2$\,fs timestep, using Hamiltonian replica exchange with an interval of $2$\,ps to enhance sampling.\cite{sug99a,swendsen86} Free energy data were analyzed using Alchemlyb and the UBAR solver inspired by Giese and York.\cite{giese2021variational}

\subsection{Torsion potential refitting with Torpenter}

Torpenter (short for ``Torsion Carpenter'') is a package for refitting the torsion potentials in a molecule according to QM or ML calculations, analogously to the \texttt{BespokeFit} and \texttt{QubeKit} packages.\cite{bespokefit,qubekit} First, the torsion profile is created by scanning the torsions. For this, the \texttt{TorsionDrive} package\cite{torsiondrive} is used in combination with \texttt{ASE}\cite{ase} using the ANI-2x neural network potential.\cite{ani2x} The 1-dimensional torsion scans utilized the step of $15$\degrees for each rotatable torsional degree of freedom, as selected according the \texttt{OpenMM Fragmenter} tool.\cite{Eastman2023a} Constraints, as implemented in the FixInternals tool of \texttt{ASE}, are applied to the selected torsion, followed by an energy minimization using the ANI-2x neural network potential in the \texttt{ASE} package. Based on the potential energies of the different rotational substates, the \texttt{ForceBalance} package\cite{forcebalance} is used to create the torsion potential with predetermined defaults. The bespoke force field parameters are then saved as an offxml file for the use in {Open Force Field}.\cite{off1}
 
\subsection{End-state Corrections}

The end-state corrections for the protein-ligand binding free energies were conducted analogously to previous solvation free energy calculations described in Ref.~\citenum{karwounopoulos2023insights}. Equilibrium simulations were conducted for both the MM and the ML levels of theory, which then served as starting points for bidirectional nonequilibrium switching simulations to calculate the free energy difference with the Crooks free energy estimator.\cite{cro00a} 

For each compound, three independent MM simulations were performed both in aqueous solution and in the bound state using OpenMM 8.1\cite{Eastman2023a} either with the Open Force Field 2.2.0 (OFF2.2.0)\cite{off2} or the Torpenter-derived force field described in the previous section. The molecules were solvated in TIP3P water,\cite{Jorgensen1983} and the chemical bonds of water were constrained with the SETTLE algorithm.\cite{Miyamoto1992}  The simulations were conducted at a temperature of $300$\,K, using Langevin dynamics with a friction constant of $1$\,ps$^{-1}$, and a pressure of $1$\,bar, using a Monte Carlo barostat.\cite{Aaqvist2004, Chow1995}  The time step was $2$\,fs,  using hydrogen mass repartitioning for all non-water hydrogens to set the hydrogen masses to $3$\,amu. The electrostatic interactions were computed with the particle mesh Ewald method,\cite{pme,Essmann1995} using a short-range cutoff of 10\,{\AA}. Before each simulation, the geometry of the solute was optimized using the L-BFGS minimizer, followed by $1$\,ns of constant volume equilibration. The  production simulation with constant pressure was performed for 5\,ns, whereof the first nanosecond was employed as equilibration. From the last 4\,ns, $300$ frames were written using a saving frequency of $12$\,ps, and used as the starting points for the NEQ switches.  

The ML/MM simulations were carried out analogously to the MM simulations, except that the intramolecular interactions of the ligand were modelled with the ANI-2x neural network potential using a mechanical embedding approach as implemented in \texttt{OpenMM-ML}.\cite{ani2x,rufa2020towards} Thus, the interactions between the ligand and the environment are still treated at the MM level. The high-performance ANI-2x implementation \texttt{NNPOPS} (v.0.6) was used.\cite{Galvelis2023NNP/MM:Mechanics}. 

$300$ nonequilibrium (NEQ) switching simulations were started from both the MM and ML/MM trajectories to calculate the free energy difference for the end-state corrections. The NEQ protocol consisted of $5$\,ps simulations with a $1$\,fs time step where the energy function is slowly interpolated from one state to the other (either MM$\rightarrow$ML, or ML$\rightarrow$MM). This was achieved by using the coupling parameter $\lambda$ to interpolate between the MM and ML/MM potential energy functions according to $U = (1-\lambda) U_{\mathrm{MM}} + \lambda U_{\mathrm{ML/MM}}$, or vice versa. The $\lambda$ variable was adjusted in an alternating sequence of $1$ perturbation and $1$ propagation time steps. In each perturbation step, the coupling parameter was adjusted to $\lambda = t/\tau$, where $t$ is the current time step and $\tau$ the total number of steps of the protocol. The nonequilibrium work value, W, is updated at each perturbation step using $W_{t} = W_{t-1} + U_{t+1}(x_{t+1}) - U_{t}(x_{t+1})$. 
Bennett's acceptance ratio, as implemented in \texttt{pymbar},\cite{Shirts2008} was used to calculate the free energy difference from the nonequilibrium work values.  

\section{Results and Discussion}
\label{sec:results}

\begin{table}[]
\caption{Comparison of the mean absolute errors (MAE) of the relative binding free energies and their corresponding standard deviations from error propagation using the Open Force Field 2.2.0 (OFF2.2.0), the refitted force field based on torsion potentials from Torpenter with ANI-2x (TOR), and the corresponding end-state corrected metrics (EC) using ANI-2x.  The last row shows the average results over all $108$ relative binding free energies. All data are in \kcal{}.}\label{tab:comparison}
\begin{tabular}{|l|l|l||l|l|}
\hline
\textbf{MAE}  & \multicolumn{2}{c||}{\textbf{MM}} & \multicolumn{2}{c|}{\textbf{ML/MM}} \\ 
\hline
      & \textbf{OFF2.2.0}  & \textbf{TOR} & \textbf{EC-OFF2.2.0}   & \textbf{EC-TOR}  \\
\hline  
TYK2    & $0.6$ $\pm$ $0.2$ & $0.7$ $\pm$ $0.2$ & $0.6$ $\pm$ $0.9$ & $0.6$ $\pm$ $0.7$\\ 
CDK2    & $0.6$ $\pm$ $0.5$ & $0.6$ $\pm$ $0.5$ & $0.6$ $\pm$ $0.7$ & $0.8$ $\pm$ $0.6$\\ 
JNK1    & $0.5$ $\pm$ $0.2$ & $0.5$ $\pm$ $0.2$ & $1.1$ $\pm$ $1.8$ & $1.0$ $\pm$ $1.2$\\ 
P38     & $1.0$ $\pm$ $0.5$ & $1.1$ $\pm$ $0.6$ & $1.0$ $\pm$ $0.7$ & $1.1$ $\pm$ $0.7$\\ 
\hline
\textbf{Average} & $\mathbf{0.8}$ $\pm$ $\mathbf{0.4}$ & $\mathbf{0.8}$ $\pm$ $\mathbf{0.4}$ & $\mathbf{0.9}$ $\pm$ $\mathbf{1.1}$ & $\mathbf{0.9}$ $\pm$ $\mathbf{0.8}$\\ 
\hline
\end{tabular}
\end{table}

\begin{figure}[]
\caption{Comparison of the $108$ relative binding free energies $\Delta\Delta G$ from MM (top) and ML/MM end-state corrections using mechanical embedding (bottom) using Open Force Field 2.2.0 (OFF2.2.0, left), and the refitted force field based on torsion potentials from Torpenter with ANI-2x (TOR, right).  The simulated $\Delta\Delta G$ values (y-axis) are plotted with respect to experiment (x-axis). The error bars correspond to the standard deviations from three repeats. The shaded area highlights deviations below $1$\,\kcal{}. Overall, the $\Delta\Delta G$ results from MM and ML/MM with mechanical embedding show a similar level of accuracy, but the ML/MM results exhibit significantly higher random fluctuations. }\label{fig:comparison}
   \begin{tabular}{cc}
    \multicolumn{2}{c}{\textbf{\underline{MM}}} \\
      (a) OFF2.2.0 & (b) TOR \\ 
      \includegraphics[width=0.47\linewidth]{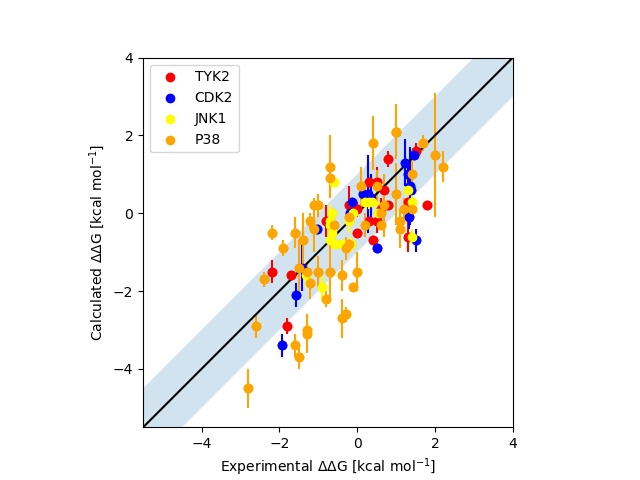}  &  \includegraphics[width=0.47\linewidth]{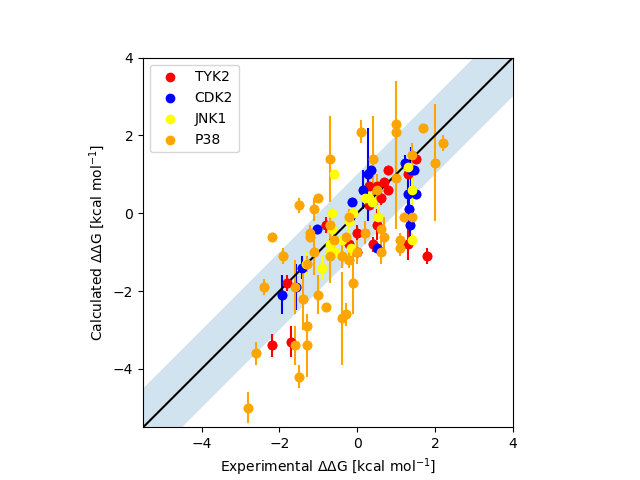}  \\
      \ \\       
    \multicolumn{2}{c}{\textbf{\underline{ML/MM}}} \\
      (c) EC-OFF2.2.0 &  (d) EC-TOR \\ 
      \includegraphics[width=0.47\linewidth]{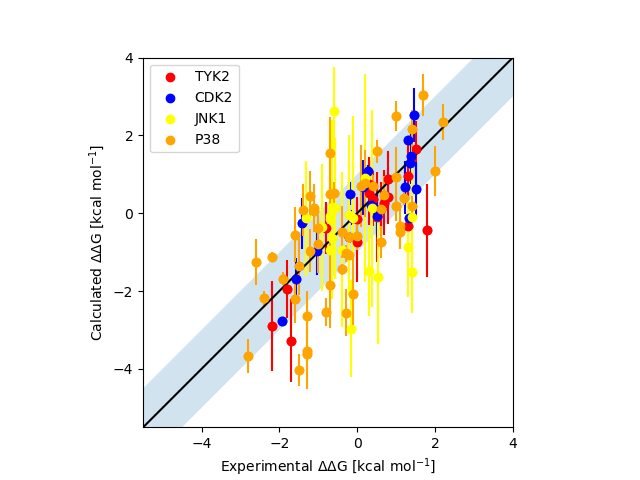}  &  \includegraphics[width=0.47\linewidth]{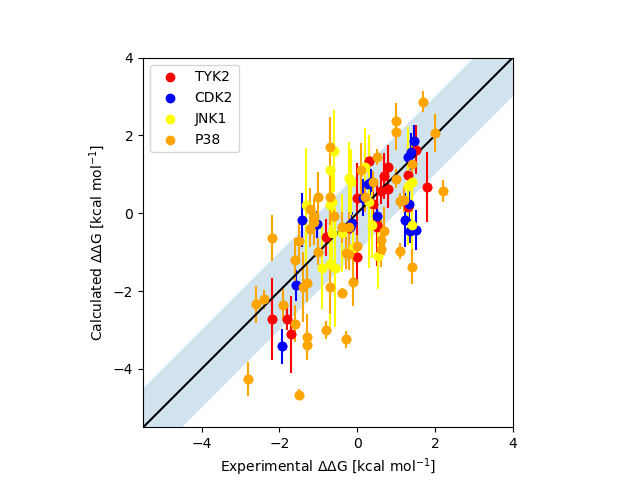}  \\      
   \end{tabular}
\end{figure}

\subsection{Verification based on reported binding affinities}

As a first step, the underlying protocol for the relative free energy simulations with OFF2.2.0 is validated based on published results. The mean absolute errors (MAE) of OFF2.2.0 with respect to experiment are reported in the first column of Table \ref{tab:comparison}. The corresponding root mean square error (RMSE) and $R^2$ data are reported in Table~S1 in the SI. 


Recently published protein-ligand affinity predictions with OFF2.0 by Hahn et al.\cite{hahn2024current} reported MAE of $1.0$\,\kcal{} for TYK2 and CDK2, and 0.7\,\kcal{} for JNK1 and P38. The MAE from OFF2.2.0 in Table~\ref{tab:comparison} are slightly better for TYK2 (0.6\,\kcal{}), CDK2 (0.6\,\kcal{}), and JNK1 (0.5\,\kcal{}), and slightly worse for P38 (1.0\,\kcal{}).  Overall, the average MAE of the OFF2.0 results by Hahn et al. for the four systems is $0.8$\,\kcal{}, which is identical with the average MAE here (0.8\,\kcal{}). Thus, the present MM relative binding free  energy results agree very well with the results of the longer $6$\,ns nonequilibrium protocol by Hahn et al.\cite{hahn2024current}.  

\subsection{Effect of torsion fitting with Torpenter}

 The mean absolute errors (MAE) with respect to experiment of OFF2.2.0 and the refitted parameters from Torpenter (TOR) are reported in the first two columns Table~\ref{tab:comparison}. Overall, OFF2.2.0 and TOR yield the same level of accuracy with MAE of $0.8$\,\kcal{}, as well as the same level of precision with average standard deviations of $0.4$\,\kcal. This is also shown by the plots of the $\Delta\Delta G$ results in Fig.~\ref{fig:comparison}a and Fig.~\ref{fig:comparison}b, where OFF2.2.0 and TOR yield very similar results. Interestingly, five data points from P38 (shown in orange) exhibit standard deviations above $1$\,\kcal{} which are markedly higher than with OFF2.2.0. This suggests that, in some cases, the ML energy surface that was used for the reparametrization of the torsion potentials might be more rugged than the MM energy surface of OFF2.2.0.     

A direct comparison of all RBFE from TOR with respect to the corresponding values from OFF2.2.0 yields a mean signed deviation of just $-0.1$\,\kcal{} and a mean absolute deviation of $0.4$\,\kcal{}. The mean unsigned deviation between TOR and OFF2.2.0 corresponds to the average standard deviation of the TOR results ($0.4$\,\kcal{}), which indicates that the observed differences are most likely the result of random fluctuations. The corresponding RMSE and $R^2$ data in Table~S1 of the Supporting Information also indicate that both methods yield the same level of accuracy and precision. This is quite remarkable, considering that the torsion potentials of the OFF2.2.0 force field were specifically fine-tuned for the compounds in the considered benchmark systems.  The results demonstrate that the Torpenter-derived torsion parameters from the fast neural network potential ANI-2x are on a par with the highly optimized parameters based on the B3LYP-D3BJ/DZVP level of theory, which suggests that Torpenter could be used for a fast on-the-fly reparametrization of all rotatable torsions of ligands in an automated drug discovery pipeline. 

\subsection{Effect of ML/MM end-state corrections with mechanical embedding}

The  two rightmost columns of Table~\ref{tab:comparison} show the mean absolute errors and standard deviations of relative binding free energies that include MM$\rightarrow$ML/MM end-state correction based mechanical embedding with ANI-2x. Because of error propagation, the standard deviations include both the uncertainty of the MM results and of the end-state corrections. The variance of the end-state corrections depends on the number and switching speed of the NEQ work calculations,\cite{dellago2013computing} and the phase-space overlap between the MM and ML/MM energy surfaces.\cite{multiconv} For example, previous free energy results based on end-state corrections with QM/MM have reported standard deviations between $0.1$ and $2.9$\,\kcal{} for hydration free energies (with  average standard deviations of about $0.6$ to $1.0$\,\kcal{}).\cite{qmmmsampl4,qmdrude} For transfer free energies between water and cyclohexane, the reported standard deviations lie between $0.4$ and $4.4$\,\kcal{}, with an average of $1.5$\,\kcal{},\cite{qmmmsampl5} and, for host-guest binding calculations, the range of the standard deviations is between $0.4$ and $1.3$\,\kcal{}.\cite{wang2018predicting}  Even when using best practices by employing NEQ switching and the Crooks estimator, the standard deviations of end-state corrections can rise up to 9.7\,\kcal{} for challenging molecules in the HiPen database.\cite{kearns2019good}  Thus, while the error bars of the MAE with ML/MM in Table~\ref{tab:comparison} are $1.2$ to $9$-times  higher than the corresponding MM results, they lie well in the range that can be expected for end-state corrections. 

Overall, the average MAE with EC from ML/MM with mechanical embedding are $0.9$\,\kcal{} for both OFF2.2.0 and TOR. Considering that the average standard deviations are $1.1$ and $0.9$\,\kcal{}, respectively, the differences between the MM and ML/MM results with mechanical embedding are not statistically significant. This view is also supported by the comparison of the MM and ML/MM results with mechanical embedding in Fig.~\ref{fig:comparison}, where no systematic differences are discernible. A direct comparison of the individual RBFE results from EC-OFF2.2.0 and OFF2.2.0 yields a mean unsigned deviation of $0.4$\,\kcal and a mean absolute deviation of $0.6$\,\kcal, which is smaller than the mean standard deviation of $1.1$\kcal{} of the EC-OFF2.2.0 results. Likewise, the mean unsigned deviation between EC-TOR and TOR is $-0.1$\,\kcal and the corresponding mean absolute deviation is $0.6$\,\kcal, which is smaller than the mean standard deviation of $0.8$\kcal{} of the EC-TOR results. Thus, any observed improvement or loss of accuracy is likely attributable to random noise. 

Focusing on RBFE for the TYK2 benchmark,  Ref.~\citenum{rufa2020towards} reported an MAE of $0.7$\,\kcal using ML/MM end-state corrections with ANI-2x. The MAE for EC with ANI-2x based on simulations with OFF2.2.0 and TOR in the first row of Table~\ref{tab:comparison} are slightly better with of $0.6$\,\kcal{}. Thus, both the MM simulations with OFF2.2.0 and TOR, and the ML/MM simulations with EC with mechanical embedding with ANI-2x and the results from Rufa et al.\ yield the same level of accuracy for the TYK2 benchmark, considering the uncertainties of the free energy results. The main difference compared to previously reported relative binding free energy calculations in Ref.\citenum{rufa2020towards} is that the previous MM simulations with OFF1.0.0 yielded an MAE of $1.0$\,\kcal{}, while MM simulations with OFF2.2.0 yield an MAE of $0.6$\,\kcal{} (first column of Table~\ref{tab:comparison}). The main differences between OFF1.0 and OFF2.0 is the refitting of the Lennard-Jones parameters,\cite{off2} which affects both the protein-ligand interactions and the intramolecular interactions of the ligand, as well as the fitting of a series of torsions to QM data.

In Ref.~\citenum{sabanes2024enhancing}, RBFE with ANI-2x and mechanical embedding were calculated with the alchemical transfer method\cite{wu2021alchemical} and compared to the GAFF2 force field. Within Ref.~\citenum{sabanes2024enhancing}, the reported MAE of relative binding free energies with ML/MM are $0.5$\,\kcal{} for TYK2, $0.7$\,\kcal{} for CDK2, $0.7$\,\kcal{} for JNK1, and $0.9$\,\kcal{} for P38. While the number of alchemical transformations for TYK2, CDK2, JNK1, and P38 in Ref.~\citenum{sabanes2024enhancing} was $84$, which is smaller than the number of transformations in the present study (108), the MAE in Table~\ref{tab:comparison} and in Ref.~\citenum{sabanes2024enhancing} are very similar, and, overall, the differences are not statistically significant. The average MAE of ML/MM for the four systems TYK2, CDK2, JNK1, and P38 from Ref.~\citenum{sabanes2024enhancing} is $0.7$\,\kcal{}, which is comparable to the results here.  The MAE of MM with the GAFF 2.11 force field from Ref.~\citenum{sabanes2024enhancing} is $1.0$\,\kcal{}, which is lower accuracy than the MAE of MM with OFF2.2.0 in Table~\ref{tab:comparison}. While the GAFF 2.11 data  suggests that mechanical embedding using ANI-2X has an average accuracy improvement of  approximately 0.3 \kcal{} over the pure MM simulations, the results in Table~\ref{tab:comparison} suggest a decrease of accuracy by approximately $0.1$\,\kcal{}. However, due to the high standard deviations of the ML/MM results, the differences are not statistically significant, and any differences might be explained by the underlying force field.

The largest deviations of the end-state corrections are observed for JNK1 using EC-OFF2.2.0, with an MAE of $1.2$ $\pm$ $1.9$\,\kcal{} (third column of Table~\ref{tab:comparison}). Only $6$ of the $21$ RBFE calculations of JNK1 exhibit errors of more than 1.5\,\kcal{}, and all of these can be traced back to just two ligands: 18634\_1 and 18628\_1 (\color{black}{Table~S4} of the SI). Ligand 18628\_1 is characterized by the addition of a methyl group to the common core, while ligand 18634\_1 involves the addition of two methoxy groups (Fig.~S11 in the SI). Given the simplicity of those additions, an error of the force field or ML potential is highly unlikely. Yet, the ML/MM end-state corrections for 18634\_1 and 18628\_1 exhibit some of the largest magnitudes in this benchmark system, with $3.8$ and $3.3$\,\kcal{}, respectively. As evidenced by the histograms of the nonequilibrium work between MM and ML/ML (Fig.~S5 in the SI),  two conformational clusters need to be sampled, which leads to high uncertainties for JNK1. While most of the ML/MM simulations show a preference for the peak on the right side, ligand 18634\_1 mostly samples the left peak. In 18628-1, the MM simulation shows an unusual extra peak on the left side, which is not present in the other MM simulations. Therefore, the high MAE of JNK1 are the result of sampling errors in the end-state corrections of those two ligands.     

The sampling of the MM simulations also influences the variance of the end-state corrections. While the overall MAE of the end-state corrections  with OFF2.2.0 exhibits a standard deviation of $1.1$\,\kcal{}, the corresponding standard deviation of the MAE with refitted torsion potentials from TOR is $0.8$\,\kcal{}. This difference of $0.3$\,\kcal{} can be explained by an increased phase space overlap between the MM and ML/MM energy surfaces in TOR, since both the torsion scans for the refitting and the end-state corrections were conducted with the ANI-2x potential. Therefore, matching the MM and ML/MM energy surfaces with a refitting procedure can also be beneficial for end-state corrections.

\section{Conclusions} 
\label{sec:conclusions}

This work represents an independent evaluation of previous reports of improved relative binding free energy results for mechanical embedding ML/MM approaches that treat the intramolecular interactions at the ML level while the protein-ligand interactions and the environment are treated at the MM level. The present ML/MM calculations using mechanical embedding and the ANI-2x potential could reproduce the same level of accuracy as reported in Refs.~\citenum{rufa2020towards} and ~\citenum{sabanes2024enhancing}, with average mean absolute errors of approximately $0.9$ $\pm$ $0.9$\,\kcal{} for the four benchmark systems TYK2, CDK2, JNK1, and P38. This corresponds to what is considered ``chemical accuracy''. The underlying MM simulations with OFF 2.2.0 also yield about the same level of accuracy as previously reported for OFF 2.0.0 by Hahn et al.\cite{hahn2024current} with an MAE of $0.8$ $\pm$ $0.4$\kcal{}. Overall, there is no statistically significant difference between the level of accuracy of the mechanical embedding ML/MM and MM  RBFE (MAE of $0.9$ and $0.8$\,\kcal{}, respectively). This can be explained by the fact that exactly the same nonbonded parameters were used to calculate the protein-ligand interactions in MM and ML/MM with mechanical embedding. This finding also agrees with a recent report of $589$ end-state corrections of hydration free energies with ML/MM using mechanical embedding, were no significant improvements were observed (MAE of $1.0$\,\kcal{} for both OFF2.0 and ML/MM with mechanical embedding and ANI-2x).\cite{karwounopoulos2023insights} Thus, any previously reported performance gains with ML/MM mechanical embedding can probably be attributed to small inadequacies of the intramolecular interactions of the ligand in the employed force fields (OFF 1.0.0 in Ref.~\citenum{rufa2020towards} and GAFF2.11 in Ref.~\citenum{sabanes2024enhancing}). However, such findings can't be extrapolated to other force fields or systems. 

One potential cause of errors in the intramolecular interactions are the  dihedral potentials. In OFF 2.2.0, dihedral potentials are determined with torsion scans of model compounds using DFT with the B3LYP-D3BJ/DZVP level of theory. Here, the Torpenter tool was used to fit the dihedral potentials of all rotatable bonds to ML with ANI-2x. This leads to exactly the same MAE as OFF2.2.0 ($0.8$ $\pm$ $0.4$\kcal{}), which supports the validity of this strategy. Refitting torsions with ML potentials instead of DFT is fast enough to be conducted on the fly, which makes it an interesting option for automated drug discovery pipelines that might explore uncommon regions of chemical space.  Compared to end-state corrections with ML/MM, refitting the torsion potentials with ML calculations is computationally less demanding and also does not lead to elevated  standard deviations. On average, the standard deviations were $0.4$\,\kcal{} with MM simulations based on Torpenter, while the corresponding end-state corrections exhibited standard deviations of about $0.8$\,\kcal{}, which is twice as high. Both options exhibit the same level of accuracy, but the increased precision of  Torpenter is likely to lead to a more robust performance, as false positives and negatives due to random fluctuations are avoided. Also, the refitting approach only requires a fraction of the computational costs of ML/MM simulations and does not require significant software changes in the drug discovery pipeline.  

While ML/MM simulations with mechanical embedding lead to the same level of accuracy as well-parameterized force fields, they are unlikely to significantly outperform MM simulations because the protein-ligand interactions are still calculated at the MM level. Going forward with ML/MM end-state corrections of protein-ligand binding, there are two main challenges that need to be addressed: First, the low precision of end-state corrections, which arises from the lack of phase space overlap between the MM and ML/MM energy surfaces. This could be addressed with more sampling, enhanced sampling techniques, or the generation of intermediate states between the MM and ML/MM energy surfaces. Using parameter refitting also improved the convergence of the end-state corrections. Second, it will be necessary to increase the accuracy of the ML/MM representation. This could be done by either employing electrostatic embedding approaches, increasing the size of the ML region, or both. Future work will also explore the use of other ML potentials for protein-ligand binding. 

\section*{Supporting Information} 
Table~S1 provides additional error metrics (RMSE, R$^2$, and Kendall $\tau$) for the four benchmark systems for the two force fields and the corresponding end-state corrections. Tables~S2 to S9 list the RBFE results for each mutation together with the experimental value and the respective end-state corrections. Figures~S1 to  S8 illustrate the work distributions for the nonequilibrium switching simulations between MM and ML/MM. Figures~S9, S10, S11, and S12 show the structures of the ligands used for each benchmark system.

\section*{Acknowledgement} 
The authors would like to thank Stefan Boresch, Marcus Wieder, and Jody Barbeau for very helpful discussions.

\setstretch{1.0}
\bibliography{bib.bib,references_jabref}

\end{document}


\maketitle

\begin{table}[H]
\caption{Comparison of the root mean square errors (RMSE), correlations coefficients (R$^2$), Kendall $\tau$ of the relative binding free energies using the OpenFF force field 2.2.0 (OFF2.2.0), the refitted force field based on torsion potentials from Torpenter with ANI-2x (TOR), and the corresponding end-state corrected metrics (EC) using ANI-2x.  The last row shows the average results over all 108 relative binding free energies. All RMSE are in \kcal{}.}\label{tab:comparison2}
\begin{tabular}{|l|lll|lll||lll|lll|}
\hline
& \multicolumn{6}{c||}{MM} & \multicolumn{6}{c|}{ML/MM} \\ 
\hline
         & \multicolumn{3}{c|}{\textbf{OFF2.2.0}} & \multicolumn{3}{c||}{\textbf{TOR}} & \multicolumn{3}{c|}{\textbf{EC-OFF2.2.0}} & \multicolumn{3}{c|}{\textbf{EC-TOR}} \\
         & RMSE    & R$^2$  & $\tau$    & RMSE       & R$^2$    & $\tau$    & RMSE     & R$^2$   & $\tau$    & RMSE       & R$^2$     & $\tau$    \\ 
\hline
TYK2     & $0.8$ & $0.6$ & $0.5$ & $1.0$ & $0.6$ & $0.5$ & $0.9$ & $0.7$ & $0.5$ & $0.7$ & $0.7$ & $0.5$  \\ 

CDK2     & $0.9$ & $0.6$ &  $0.5$ & $0.8$ & $0.6$ & $0.6$ & $0.7$ & $0.7$ & $0.6$ & $1.0$ & $0.5$ &  $0.4$\\   

JNK1     & $0.7$ & $0.3$ & $0.4$ & $0.7$ & $0.4$ & $0.5$ & $1.5$ & $0.1$ & $-0.1$& $1.1$ & $0.0$ & $0.1$ \\ 

P38      & $1.2$ & $0.5$ & $0.5$ & $1.3$ & $0.5$ & $0.5$ & $1.2$ & $0.5$ & $0.5$ & $1.3$ & $0.4$ & $0.5$ \\ 
\hline
\textbf{Average}  & $\mathbf{1.0}$  & $\mathbf{0.5}$ & $\mathbf{0.5}$ & $\mathbf{1.1}$ & $\mathbf{0.5}$ & $\mathbf{0.5}$ & $\mathbf{1.1}$ & $\mathbf{0.4}$ & $\mathbf{0.5}$ & $\mathbf{1.1}$ & $\mathbf{0.4}$ & $\mathbf{0.4}$ \\ 
\hline
\end{tabular}
\end{table}

\begin{table}[H]
    \centering
    \caption{TYK2 relative binding free energy results, based on MM simulations with Open Force Field (OFF2.2.0), reparameterized torsion potentials with Torpenter (TOR) and end-state corrections (EC) based on mechanical embedding with ANI-2x.  The error bars show the respective standard deviations. The experimental reference results are denoted with EXP. All data are in \kcal{}.}
    \resizebox{\textwidth}{!}{ 
    \begin{tabular}{@{}lcr|rr|rr@{}}
        \hline
        Ligand A & Ligand B & EXP & OFF2.2.0 & EC-OFF2.2.0 & TOR & EC-TOR \\ \hline
        ejm\_31 & ejm\_45 & 0 & $0.1 \pm 0.2$ & $-0.7 \pm 1.1$ & $-0.5 \pm 0.2$ & $0.4 \pm 0.9$ \\
        ejm\_31 & ejm\_47 & -0.2 & $0.2 \pm 0.5$ & $-0.6 \pm 1.1$ & $-0.8 \pm 0.2$ & $-0.4 \pm 1.0$ \\
        ejm\_31 & ejm\_48 & 0.5 & $0.8 \pm 0.4$ & $-0.1 \pm 1.2$ & $-0.3 \pm 0.4$ & $-0.3 \pm 1.1$ \\
        ejm\_31 & ejm\_49 & 1.8 & $0.2 \pm 0.1$ & $-0.4 \pm 1.2$ & $-1.1 \pm 0.2$ & $0.7 \pm 0.9$ \\
        ejm\_31 & jmc\_23 & -2.2 & $-1.5 \pm 0.3$ & $-2.9 \pm 1.2$ & $-3.4 \pm 0.3$ & $-2.7 \pm 1.1$ \\
        ejm\_31 & jmc\_27 & -1.7 & $-1.6 \pm 0.1$ & $-3.3 \pm 1.1$ & $-3.3 \pm 0.4$ & $-3.1 \pm 1.1$ \\
        ejm\_42 & ejm\_43 & 1.5 & $1.6 \pm 0.2$ & $1.6 \pm 0.8$ & $1.4 \pm 0.1$ & $1.6 \pm 0.6$ \\
        ejm\_42 & ejm\_50 & 0.8 & $0.2 \pm 0.1$ & $0.4 \pm 0.7$ & $0.6 \pm 0.1$ & $1.2 \pm 0.6$ \\
        ejm\_42 & ejm\_54 & -0.8 & $-0.2 \pm 0.4$ & $-0.4 \pm 0.8$ & $-0.3 \pm 0.2$ & $-0.6 \pm 0.5$ \\
        ejm\_43 & ejm\_44 & 0.8 & $1.4 \pm 0.2$ & $0.9 \pm 0.7$ & $1.1 \pm 0.1$ & $0.6 \pm 0.5$ \\
        ejm\_44 & ejm\_55 & -1.8 & $-2.9 \pm 0.2$ & $-2 \pm 0.8$ & $-1.8 \pm 0.2$ & $-2.7 \pm 0.4$ \\
        ejm\_45 & ejm\_50 & 0.6 & $0.1 \pm 0.3$ & $0.1 \pm 0.8$ & $0.4 \pm 0.2$ & $0.6 \pm 0.4$ \\
        ejm\_46 & jmc\_27 & 0 & $-0.5 \pm 0.1$ & $-0.2 \pm 0.6$ & $-1.0 \pm 0.1$ & $-1.1 \pm 0.7$ \\
        ejm\_46 & jmc\_30 & 0.4 & $-0.7 \pm 0.1$ & $0.4 \pm 0.7$ & $-0.8 \pm 0.2$ & $0.2 \pm 0.6$ \\
        ejm\_47 & ejm\_55 & 0.5 & $-0.2 \pm 0.3$ & $0 \pm 0.8$ & $0.7 \pm 0.1$ & $-0.4 \pm 0.5$ \\
        ejm\_48 & ejm\_49 & 1.3 & $-0.6 \pm 0.4$ & $-0.3 \pm 1.1$ & $-0.8 \pm 0.4$ & $1.0 \pm 0.8$ \\
        ejm\_54 & ejm\_55 & 1.3 & $0.3 \pm 0.1$ & $1.0 \pm 0.8$ & $1.0 \pm 0.1$ & $0.2 \pm 0.1$ \\
        jmc\_23 & jmc\_28 & 0.7 & $0.6 \pm 0.1$ & $0.3 \pm 0.8$ & $0.8 \pm 0.1$ & $1.0 \pm 0.6$ \\
        jmc\_27 & jmc\_28 & 0.3 & $0.8 \pm 0.1$ & $0.8 \pm 0.7$ & $0.7 \pm 0.0$ & $1.4 \pm 0.5$ \\
        jmc\_27 & jmc\_30 & 0.3 & $-0.2 \pm 0.1$ & $0.5 \pm 0.8$ & $0.2 \pm 0.1$ & $1.3 \pm 0.5$ \\\hline
        \end{tabular}
    } 
    \label{tab:rbfe_tyk2}
\end{table}

\begin{table}[H]
    \centering
    \caption{CDK2 relative binding free energy results, based on MM simulations with Open Force Field (OFF2.2.0), reparameterized torsion potentials with Torpenter (TOR) and end-state corrections (EC) based on mechanical embedding with ANI-2x.  The error bars show the respective standard deviations. The experimental reference results are denoted with EXP. All data are in \kcal{}.}
    \resizebox{\textwidth}{!}{ 
    \begin{tabular}{@{}lcr|rr|rr@{}}
        \hline
        Ligand A & Ligand B & EXP & OFF2.2.0 & EC-OFF2.2.0 & TOR & EC-TOR \\ \hline
        1h1q & 1h1r & 0.51 & $-0.9 \pm 0.1$ & $-0.1 \pm 0.6$ & $-0.9 \pm 0.1$ & $-0.1 \pm 0.2$ \\
        1h1q & 1oi9 & -1.56 & $-2.1 \pm 0.3$ & $-1.7 \pm 0.6$ & $-1.9 \pm 0.6$ & $-1.9 \pm 0.7$ \\
        1h1r & 1oiu & -1.42 & $-1.4 \pm 0.6$ & $-0.3 \pm 0.9$ & $-1.4 \pm 0.3$ & $-0.2 \pm 0.8$ \\
        1h1r & 20 & -1.03 & $-0.4 \pm 0.1$ & $-1.0 \pm 0.6$ & $-0.4 \pm 0.1$ & $-0.3 \pm 0.4$ \\
        1h1r & 21 & -0.14 & $0.3 \pm 0.1$ & $-0.1 \pm 0.2$ & $0.3 \pm 0.1$ & $-0.3 \pm 0.3$ \\
        1h1r & 22 & -0.18 & $0 \pm 0.3$ & $0.5 \pm 0.4$ & $-0.1 \pm 0.1$ & $-0.4 \pm 0.2$ \\
        1h1s & 28 & 0.15 & $0.5 \pm 0.2$ & $0.7 \pm 0.7$ & $0.6 \pm 0.5$ & $0.4 \pm 0.7$ \\
        1h1s & 29 & 1.38 & $0.6 \pm 0.4$ & $1.5 \pm 0.8$ & $1.1 \pm 0.6$ & $1.6 \pm 0.8$ \\
        1h1s & 30 & 1.45 & $1.5 \pm 0.1$ & $2.5 \pm 0.7$ & $1.1 \pm 0.1$ & $1.9 \pm 0.4$ \\
        1h1s & 32 & 1.51 & $-0.7 \pm 0.3$ & $0.6 \pm 0.8$ & $0.5 \pm 0.1$ & $-0.4 \pm 0.5$ \\
        1oi9 & 26 & 1.32 & $-0.1 \pm 0.3$ & $-0.1 \pm 0.3$ & $0.1 \pm 0.5$ & $0.2 \pm 0.7$ \\
        1oiu & 22 & 1.24 & $1.3 \pm 0.6$ & $0.7 \pm 0.9$ & $1.3 \pm 0.2$ & $-0.2 \pm 0.7$ \\
        1oiy & 26 & 1.37 & $0.7 \pm 1$ & $1.3 \pm 1$ & $-0.3 \pm 0.4$ & $-0.5 \pm 0.5$ \\
        1oiy & 31 & 0.27 & $0.5 \pm 1$ & $1.1 \pm 1$ & $1.0 \pm 1.2$ & $0.8 \pm 1.2$ \\
        21 & 32 & -1.93 & $-3.4 \pm 0.3$ & $-2.8 \pm 0.3$ & $-2.1 \pm 0.5$ & $-3.4 \pm 0.7$ \\
        28 & 30 & 1.3 & $1.0 \pm 0.3$ & $1.9 \pm 0.3$ & $0.5 \pm 0.4$ & $1.4 \pm 0.6$ \\
        29 & 31 & 0.35 & $0.4 \pm 0.6$ & $0.2 \pm 0.6$ & $1.1 \pm 0.1$ & $0.8 \pm 0.4$ \\\hline
        \end{tabular}
    } 
    \label{tab:rbfe_cdk2}
\end{table}

\begin{table}[H]
    \centering
    \caption{JNK1 relative binding free energy results, based on MM simulations with Open Force Field (OFF2.2.0), reparameterized torsion potentials with Torpenter (TOR) and end-state corrections (EC) based on mechanical embedding with ANI-2x.  The error bars show the respective standard deviations. The experimental reference results are denoted with EXP. All data are in \kcal{}.}
    \resizebox{\textwidth}{!}{ 
    \begin{tabular}{@{}lcr|rr|rr@{}}
        \hline
        Ligand A & Ligand B & EXP & OFF2.2.0 & EC-OFF2.2.0 & TOR & EC-TOR \\ \hline
        18624 & 18626 & -0.4 & $-0.8 \pm0.0$ & $-0.9 \pm 2$ & $-0.7 \pm 0.0$ & $-0.5 \pm 1.0$ \\
        18624 & 18630 & -0.65 & $-0.5 \pm 0.1$ & $-0.6 \pm 1.6$ & $-0.4 \pm 0.1$ & $-0.5 \pm 0.8$ \\
        18624 & 18633 & -0.69 & $-0.7 \pm0.0$ & $-1.0 \pm 1.2$ & $-0.8 \pm0.0$ & $0.2 \pm 0.9$ \\
        18625 & 18628 & -0.6 & $0.8 \pm 0.1$ & $2.6 \pm 1.1$ & $1.0 \pm 0.1$ & $1.5 \pm 1.1$ \\
        18625 & 18631 & -1.32 & $-1.6 \pm 0.3$ & $-0.1 \pm 1.5$ & $-1.3 \pm 0.3$ & $0.2 \pm 1.5$ \\
        18626 & 18627 & 0.39 & $0.3 \pm 0.1$ & $0.1 \pm 2.5$ & $0.3 \pm 0.1$ & $-0.4 \pm 0.9$ \\
        18626 & 18629 & 0.19 & $0.3 \pm 0.2$ & $0.9 \pm 2.7$ & $0.4 \pm 0.2$ & $1.1 \pm 1.1$ \\
        18626 & 18632 & -0.22 & $-0.2 \pm 0.3$ & $-0.1 \pm 2.1$ & $-0.2 \pm 0.3$ & $0.9 \pm 1$ \\
        18627 & 18630 & -0.66 & $0.0\pm0.0$ & $0.2 \pm 2.2$ & $0.0\pm0.0$ & $0.4 \pm 0.7$ \\
        18627 & 18633 & -0.7 & $-0.2 \pm0.0$ & $-0.1 \pm 2$ & $-0.3 \pm0.0$ & $1.2 \pm 0.8$ \\
        18628 & 18635 & 1.42 & $-0.6 \pm 0.2$ & $-1.5 \pm 1.1$ & $-0.7 \pm 0.2$ & $-0.3 \pm 0.9$ \\
        18629 & 18635 & 1.4 & $0.3 \pm 0.4$ & $-0.1 \pm 2.1$ & $0.6 \pm 0.4$ & $0.8 \pm 1.1$ \\
        18631 & 18634 & -0.58 & $-0.8 \pm 0.1$ & $0.1 \pm 1.8$ & $-1.0 \pm 0.1$ & $-1.4 \pm 1.5$ \\
        18632 & 18634 & -0.91 & $-1.9 \pm 0.3$ & $-0.4 \pm 1.7$ & $-1.4 \pm 0.3$ & $-1.4 \pm 1.1$ \\
        18634 & 18637 & -0.15 & $-0.8 \pm 0.2$ & $-3 \pm 1.3$ & $-0.9 \pm 0.2$ & $0.8 \pm 0.9$ \\
        18634 & 18638 & -0.1 & $0.0\pm 0.2$ & $-0.1 \pm 2.6$ & $0.0\pm 0.2$ & $-1.0 \pm 0.9$ \\
        18634 & 18652 & -0.69 & $-0.3 \pm 0.5$ & $-0.1 \pm 1.9$ & $-1.0 \pm 0.5$ & $-1.4 \pm 1.7$ \\
        18634 & 18658 & 0.3 & $0.3 \pm 0.2$ & $-1.5 \pm 1.2$ & $0.4 \pm 0.2$ & $0.2 \pm 1.7$ \\
        18634 & 18659 & 0.53 & $0.0\pm 0.3$ & $-1.6 \pm 1.8$ & $-0.1 \pm 0.3$ & $-1.1 \pm 0.9$ \\
        18634 & 18660 & 1.3 & $0.6 \pm0.0$ & $-0.9 \pm 1.3$ & $1.2 \pm0.0$ & $0.7 \pm 1.6$ \\
        18637 & 18639 & 0.41 & $0.6 \pm 0.3$ & $1.6 \pm 1.7$ & $0.2 \pm 0.3$ & $-2.4 \pm 1.1$ \\
        18638 & 18639 & 0.36 & $-0.2 \pm 0.3$ & $-1.3 \pm 2.8$ & $-0.6 \pm 0.3$ & $-0.6 \pm 1.1$ \\
        18639 & 18652 & -0.95 & $-0.1 \pm 0.7$ & $1.3 \pm 2.3$ & $-0.3 \pm 0.7$ & $0.3 \pm 1.9$ \\
        18639 & 18660 & 1.04 & $0.8 \pm 0.3$ & $0.6 \pm 1.7$ & $1.8 \pm 0.3$ & $2.3 \pm 1.7$ \\
        18658 & 18659 & 0.23 & $-0.3 \pm 0.2$ & $-0.1 \pm 1.4$ & $-0.5 \pm 0.2$ & $-1.3 \pm 1.5$ \\\hline
        \end{tabular}
    } 
    \label{tab:rbfe_jnk1}
\end{table}

\begin{table}[H]
    \centering
    \caption{P38 relative binding free energy results, based on MM simulations with Open Force Field (OFF2.2.0), reparameterized torsion potentials with Torpenter (TOR) and end-state corrections (EC) based on mechanical embedding with ANI-2x.  The error bars show the respective standard deviations. The experimental reference results are denoted with EXP. All data are in \kcal{}.}
    \resizebox{0.7\textwidth}{!}{ 
    \begin{tabular}{@{}lcr|rr|rr@{}}
        \hline
        Ligand A & Ligand B & EXP & OFF2.2.0 & EC-OFF2.2.0 & TOR & EC-TOR \\ \hline
        2aa & 2z & -1.1 & $0.2 \pm 0.2$ & $0.1 \pm 0.6$ & $0.1 \pm 0.3$ & $-0.2 \pm 0.7$ \\
        2aa & 3flq & -0.7 & $1.2 \pm 0.8$ & $1.5 \pm 1.2$ & $-1.1 \pm 0.7$ & $0.4 \pm 1.0$ \\
        2aa & 3flw & -1.4 & $-0.7 \pm 0.7$ & $0.1 \pm 1.0$ & $-2.2 \pm 0.8$ & $-1.9 \pm 1.2$ \\
        2aa & 3fmh & -1.6 & $-0.5 \pm 0.4$ & $-0.6 \pm 0.8$ & $-1.9 \pm 0.7$ & $-1.2 \pm 0.9$ \\
        2aa & 3fmk & -2.6 & $-2.9 \pm 0.3$ & $-1.3 \pm 0.7$ & $-3.6 \pm 0.3$ & $-2.3 \pm 0.6$ \\
        2bb & 2v & 0.1 & $0.7 \pm 0.5$ & $0.7 \pm 1.2$ & $2.1 \pm 0.3$ & $1.1 \pm 0.8$ \\
        2bb & 2y & -0.7 & $-1.5 \pm 0.7$ & $-1.9 \pm 1.3$ & $-0.3 \pm 0.3$ & $-1.9 \pm 0.8$ \\
        2c & 2h & 1.0 & $2.1 \pm 0.7$ & $0.2 \pm 0.8$ & $2.1 \pm 1.3$ & $0.9 \pm 1.3$ \\
        2c & 2i & 0.4 & $1.8 \pm 0.7$ & $0.7 \pm 0.8$ & $1.4 \pm 1.1$ & $0.8 \pm 1.1$ \\
        2e & 3fln & -0.2 & $-0.8 \pm 0.1$ & $-1.1 \pm 0.2$ & $-1.2 \pm 0.2$ & $-1.0 \pm 0.5$ \\
        2e & 3flz & 1.2 & $0.1 \pm 0.3$ & $0.4 \pm 0.4$ & $-0.1 \pm 0.0$ & $0.4 \pm 0.3$ \\
        2ee & 2j & 2.2 & $1.2 \pm 0.4$ & $2.4 \pm 0.6$ & $1.8 \pm 0.2$ & $0.6 \pm 0.3$ \\
        2ee & 3fln & 1.4 & $0.1 \pm 0.1$ & $0.2 \pm 0.3$ & $-0.1 \pm 0.2$ & $-1.4 \pm 0.5$ \\
        2f & 2g & -2.2 & $-0.5 \pm 0.2$ & $-1.1 \pm 0.2$ & $-0.6 \pm 0.0$ & $-0.6 \pm 0.6$ \\
        2f & 3flz & -1.0 & $0.2 \pm 0.3$ & $-0.4 \pm 0.4$ & $0.4 \pm 0.1$ & $0.4 \pm 0.6$ \\
        2ff & 2j & 1.4 & $1.0 \pm 0.4$ & $2.2 \pm 0.6$ & $1.5 \pm 0.3$ & $1.3 \pm 0.7$ \\
        2ff & 3fln & 0.6 & $0.0 \pm 0.1$ & $0.1 \pm 0.3$ & $-0.4 \pm 0.1$ & $-0.7 \pm 0.7$ \\
        2g & 3fln & -0.2 & $-0.1 \pm 0.1$ & $-0.6 \pm 0.2$ & $-0.1 \pm 0.2$ & $-0.3 \pm 0.4$ \\
        2gg & 2m & -0.3 & $-2.6 \pm 0.2$ & $-2.6 \pm 0.6$ & $-2.6 \pm 0.3$ & $-3.2 \pm 0.4$ \\
        2gg & 2n & -1.2 & $-1.8 \pm 0.4$ & $-1.0 \pm 0.7$ & $-0.5 \pm 0.2$ & $0.1 \pm 0.6$ \\
        2gg & 2o & 0.2 & $-0.3 \pm 0.3$ & $0.8 \pm 0.9$ & $-0.5 \pm 0.3$ & $0.4 \pm 0.5$ \\
        2gg & 2r & -0.1 & $-1.9 \pm 0.1$ & $-2.1 \pm 1.0$ & $-1.8 \pm 0.8$ & $-1.8 \pm 1.0$ \\
        2gg & 2s & -0.4 & $-1.6 \pm 0.4$ & $-0.5 \pm 0.8$ & $-1.1 \pm 0.3$ & $-0.4 \pm 0.5$ \\
        2gg & 2u & -1.6 & $-3.4 \pm 0.3$ & $-2.2 \pm 0.7$ & $-3.4 \pm 0.5$ & $-2.8 \pm 0.7$ \\
        2gg & 2v & 1.7 & $1.8 \pm 0.2$ & $3.1 \pm 0.6$ & $2.2 \pm 0.1$ & $2.9 \pm 0.3$ \\
        2h & 2i & -0.6 & $-0.3 \pm 0.4$ & $0.5 \pm 0.5$ & $-0.7 \pm 0.2$ & $-0.1 \pm 0.3$ \\
        2h & 2m & -1.5 & $-3.7 \pm 0.3$ & $-4.0 \pm 0.5$ & $-4.2 \pm 0.3$ & $-4.7 \pm 0.3$ \\
        2h & 2o & -1.0 & $-1.5 \pm 0.4$ & $-0.8 \pm 0.8$ & $-2.1 \pm 0.5$ & $-1.0 \pm 0.6$ \\
        2h & 2r & -1.3 & $-3.1 \pm 0.5$ & $-3.6 \pm 1.0$ & $-3.4 \pm 0.8$ & $-3.2 \pm 1.0$ \\
        2h & 2u & -2.8 & $-4.5 \pm 0.5$ & $-3.7 \pm 0.7$ & $-5.0 \pm 0.4$ & $-4.3 \pm 0.6$ \\
        2h & 2v & 0.5 & $0.7 \pm 0.4$ & $1.6 \pm 0.5$ & $0.6 \pm 0.4$ & $1.4 \pm 0.5$ \\
        2j & 2v & 1.1 & $-0.2 \pm 0.1$ & $-0.5 \pm 0.4$ & $-0.9 \pm 0.2$ & $0.3 \pm 0.3$ \\
        2k & 2t & -1.3 & $-3.0 \pm 0.3$ & $-3.5 \pm 0.5$ & $-2.9 \pm 0.1$ & $-3.4 \pm 0.4$ \\
        2k & 3fln & -0.3 & $-0.9 \pm 0.3$ & $-1.0 \pm 0.4$ & $-0.6 \pm 0.1$ & $-1.0 \pm 0.4$ \\
        2l & 2p & 1.1 & $-0.4 \pm 0.5$ & $-0.3 \pm 0.8$ & $-0.7 \pm 0.2$ & $-1.0 \pm 0.3$ \\
        2l & 2q & 0.0 & $-1.5 \pm 0.5$ & $-0.6 \pm 0.6$ & $-1.0 \pm 0.3$ & $-0.9 \pm 0.4$ \\
        2n & 2s & 0.7 & $0.2 \pm 0.8$ & $0.5 \pm 0.9$ & $-0.6 \pm 0.5$ & $-0.5 \pm 0.8$ \\
        2p & 2x & 1.0 & $0.5 \pm 0.8$ & $0.9 \pm 1.1$ & $1.3 \pm 2.1$ & $1.4 \pm 1.4$ \\
        2q & 2x & 2.0 & $1.5 \pm 1.6$ & $1.1 \pm 1.7$ & $1.5 \pm 2.1$ & $1.6 \pm 1.6$ \\
        2t & 3fln & 1.0 & $2.1 \pm 0.4$ & $2.5 \pm 0.6$ & $2.3 \pm 0.1$ & $2.4 \pm 0.5$ \\
        2v & 2x & -1.3 & $-1.5 \pm 0.2$ & $-2.6 \pm 0.7$ & $-1.3 \pm 0.2$ & $-1.8 \pm 0.5$ \\
        2v & 2y & -0.8 & $-2.2 \pm 0.2$ & $-2.6 \pm 0.4$ & $-2.4 \pm 0.1$ & $-3.0 \pm 0.3$ \\
        2v & 3fln & -1.9 & $-0.9 \pm 0.2$ & $-1.7 \pm 0.3$ & $-1.1 \pm 0.2$ & $-2.4 \pm 0.5$ \\
        2v & 3fly & -2.4 & $-1.7 \pm 0.2$ & $-2.2 \pm 0.3$ & $-1.9 \pm 0.2$ & $-2.2 \pm 0.3$ \\
        2x & 3fly & -1.2 & $-0.2 \pm 0.2$ & $0.5 \pm 0.7$ & $-0.6 \pm 0.1$ & $-0.4 \pm 0.5$ \\
        2z & 3fly & -1.1 & $-0.4 \pm 0.6$ & $0.1 \pm 0.6$ & $-1.0 \pm 0.6$ & $-0.1 \pm 0.7$ \\
        3flq & 3fly & -1.5 & $-1.4 \pm 0.7$ & $-1.4 \pm 1.0$ & $0.2 \pm 0.2$ & $-0.7 \pm 0.6$ \\
        3flw & 3fly & -0.7 & $0.9 \pm 0.5$ & $0.5 \pm 0.6$ & $1.4 \pm 1.1$ & $1.7 \pm 1.3$ \\
        3fly & 3fmh & 0.6 & $-0.3 \pm 0.1$ & $-0.7 \pm 0.4$ & $-1.0 \pm 0.3$ & $-0.9 \pm 0.4$ \\
        3fly & 3fmk & -0.4 & $-2.7 \pm 0.5$ & $-1.4 \pm 0.5$ & $-2.7 \pm 1.2$ & $-2.0 \pm 1.2$ \\\hline
        \end{tabular}
    } 
    \label{tab:rbfe_p38}
\end{table}

\begin{table}[H]
\centering
\caption{End-state correction (EC) results and the corresponding standard deviations (STD) from three independent runs for the \textbf{TYK2} system using both the Open force field 2.2.0 (OFF2.2.0) and Torpenter (TOR).}
\begin{tabular}{l|cc|cc}
\hline
Ligand & EC-OFF2.2.0 & STD-OFF2.2.0 & EC-TOR & STD-TOR \\
\hline
ejm\_31 & 3.3 & 0.9 & 1.1 & 0.9 \\
ejm\_42 & 2.2 & 0.4 & 1.6 & 0.5 \\
ejm\_43 & 2.3 & 0.6 & 1.8 & 0.4 \\
ejm\_44 & 1.7 & 0.4 & 1.3 & 0.3 \\
ejm\_45 & 2.4 & 0.4 & 2.0 & 0.1 \\
ejm\_46 & 1.2 & 0.3 & 1.4 & 0.5 \\
ejm\_47 & 2.4 & 0.4 & 1.5 & 0.5 \\
ejm\_48 & 2.4 & 0.7 & 1.1 & 0.6 \\
ejm\_49 & 2.6 & 0.7 & 2.9 & 0.2 \\
ejm\_50 & 2.4 & 0.6 & 2.2 & 0.3 \\
ejm\_54 & 2.0 & 0.5 & 1.3 & 0.1 \\
ejm\_55 & 2.7 & 0.6 & 0.4 & 0.0 \\
jmc\_23 & 1.9 & 0.7 & 1.8 & 0.6 \\
jmc\_27 & 1.6 & 0.5 & 1.3 & 0.4 \\
jmc\_28 & 1.5 & 0.5 & 1.9 & 0.1 \\
jmc\_30 & 2.3 & 0.7 & 2.4 & 0.2 \\
\hline
average & & 0.6 & & 0.4 \\
\hline
\end{tabular}
\label{tableS:ecTYK2}
\end{table}

\begin{table}[H]
\centering
\caption{End-state correction (EC) results and the corresponding standard deviations (STD) from three independent runs for the \textbf{CDK2} system using both the Open force field 2.2.0 (OFF2.2.0) and Torpenter (TOR).}
\begin{tabular}{l|cc|cc}
\hline
Ligand & EC-OFF2.2.0 & STD-OFF2.2.0 & EC-TOR & STD-TOR \\
\hline
1h1q & -2.0 & 0.5 & -1.7 & 0.1 \\
1h1r & -1.1 & 0.2 & -0.8 & 0.1 \\
1h1s & -2.2 & 0.7 & -1.8 & 0.4 \\
1oi9 & -1.6 & 0.0 & -1.6 & 0.4 \\
1oiu & 0.0 & 0.6 & 0.4 & 0.7 \\
1oiy & -2.2 & 0.1 & -1.3 & 0.1 \\
20 & -1.7 & 0.6 & -0.7 & 0.3 \\
21 & -1.5 & 0.1 & -1.4 & 0.2 \\
22 & -0.6 & 0.2 & -1.1 & 0.1 \\
26 & -1.6 & 0.0 & -1.5 & 0.3 \\
28 & -2.1 & 0.1 & -2.0 & 0.3 \\
29 & -1.4 & 0.1 & -1.3 & 0.3 \\
30 & -1.2 & 0.0 & -1.0 & 0.2 \\
31 & -1.6 & 0.1 & -1.6 & 0.1 \\
32 & -0.9 & 0.0 & -2.7 & 0.4 \\\hline
average &    & 0.2 &   & 0.3 \\
\hline
\end{tabular}
\label{tableS:ecCDK2}
\end{table}

\begin{table}[H]
\centering
\caption{End-state correction (EC) results and the corresponding standard deviations (STD) from three independent runs for the \textbf{JNK1} system using both the Open force field 2.2.0 (OFF2.2.0) and Torpenter (TOR).}
\begin{tabular}{l|cc|cc}
\hline
Ligand & EC-OFF2.2.0 & STD-OFF2.2.0 & EC-TOR & STD-TOR \\
\hline
18624 & 2.3 & 1.1 & 2.6 & 0.8 \\
18625 & 1.4 & 0.4 & 2.8 & 0.7 \\
18626 & 2.1 & 1.7 & 2.8 & 0.6 \\
18627 & 2.0 & 1.9 & 2.2 & 0.6 \\
18628 & 3.3 & 1.0 & 3.4 & 0.8 \\
18629 & 2.7 & 2.1 & 3.6 & 0.8 \\
18630 & 2.1 & 1.1 & 2.6 & 0.3 \\
18631 & 2.9 & 1.4 & 4.3 & 1.3 \\
18632 & 2.3 & 1.2 & 3.9 & 0.7 \\
18633 & 2.0 & 0.6 & 3.6 & 0.5 \\
18634 & 3.8 & 1.1 & 4.0 & 0.8 \\
18635 & 2.3 & 0.1 & 3.8 & 0.5 \\
18637 & 1.6 & 0.6 & 5.6 & 0.3 \\
18638 & 3.7 & 2.3 & 3.0 & 0.3 \\
18639 & 2.6 & 1.6 & 3.0 & 1.0 \\
18652 & 4.0 & 1.5 & 3.6 & 1.4 \\
18658 & 2.0 & 0.2 & 3.8 & 1.5 \\
18659 & 2.2 & 1.3 & 3.0 & 0.3 \\
18660 & 2.4 & 0.7 & 3.5 & 1.3 \\
\hline
average & & 1.2 & & 0.8 \\
\hline
\end{tabular}
\label{tableS:ecJNK1}
\end{table}

\begin{table}[H]
\centering
\caption{End-state correction (EC) results and the corresponding standard deviations (STD) from three independent runs for the \textbf{P38} system using both the Open force field 2.2.0 (OFF2.2.0) and Torpenter (TOR).}
\begin{tabular}{l|cc|cc}
\hline
Ligand & EC-OFF2.2.0 & STD-OFF2.2.0 & EC-TOR & STD-TOR \\
\hline
2aa & -2.1 & 0.6 & -1.5 & 0.5 \\
2bb & -1.3 & 1.0 & 0.4 & 0.7 \\
2c & -0.2 & 0.3 & -0.2 & 0.2 \\
2e & -1.8 & 0.1 & -2.0 & 0.2 \\
2ee & -2.1 & 0.2 & -0.6 & 0.3 \\
2f & -0.9 & 0.1 & -1.4 & 0.4 \\
2ff & -2.1 & 0.3 & -1.6 & 0.6 \\
2g & -1.5 & 0.1 & -1.6 & 0.1 \\
2gg & -2.5 & 0.5 & -1.3 & 0.2 \\
2h & -2.1 & 0.3 & -1.4 & 0.1 \\
2i & -1.3 & 0.1 & -0.8 & 0.1 \\
2j & -1.0 & 0.4 & -1.8 & 0.1 \\
2k & -1.9 & 0.2 & -1.4 & 0.2 \\
2l & -2.9 & 0.3 & -2.0 & 0.2 \\
2m & -2.5 & 0.3 & -1.9 & 0.1 \\
2n & -1.7 & 0.2 & -0.7 & 0.5 \\
2o & -1.4 & 0.7 & -0.3 & 0.3 \\
2p & -2.8 & 0.5 & -2.3 & 0.1 \\
2q & -2.0 & 0.1 & -1.9 & 0.2 \\
2r & -2.7 & 0.9 & -1.2 & 0.6 \\
2s & -1.4 & 0.4 & -0.5 & 0.3 \\
2t & -2.4 & 0.4 & -1.9 & 0.3 \\
2u & -1.3 & 0.3 & -0.7 & 0.4 \\
2v & -1.3 & 0.1 & -0.6 & 0.2 \\
2x & -2.4 & 0.6 & -1.1 & 0.4 \\
2y & -1.6 & 0.3 & -1.2 & 0.1 \\
2z & -2.2 & 0.2 & -1.8 & 0.4 \\
3fln & -2.0 & 0.1 & -1.9 & 0.4 \\
3flq & -1.8 & 0.7 & 0.0 & 0.5 \\
3flw & -1.3 & 0.3 & -1.2 & 0.8 \\
3fly & -1.7 & 0.1 & -0.9 & 0.1 \\
3flz & -1.5 & 0.2 & -1.6 & 0.2 \\
3fmh & -2.2 & 0.4 & -0.8 & 0.3 \\
3fmk & -0.5 & 0.1 & -0.3 & 0.0 \\
\hline
average &  & 0.3 &  & 0.3 \\
\hline
\end{tabular}
\label{tableS:ecP38}
\end{table}

\clearpage
\begin{figure}[H]
    \centering
    \includegraphics[width=1\linewidth]{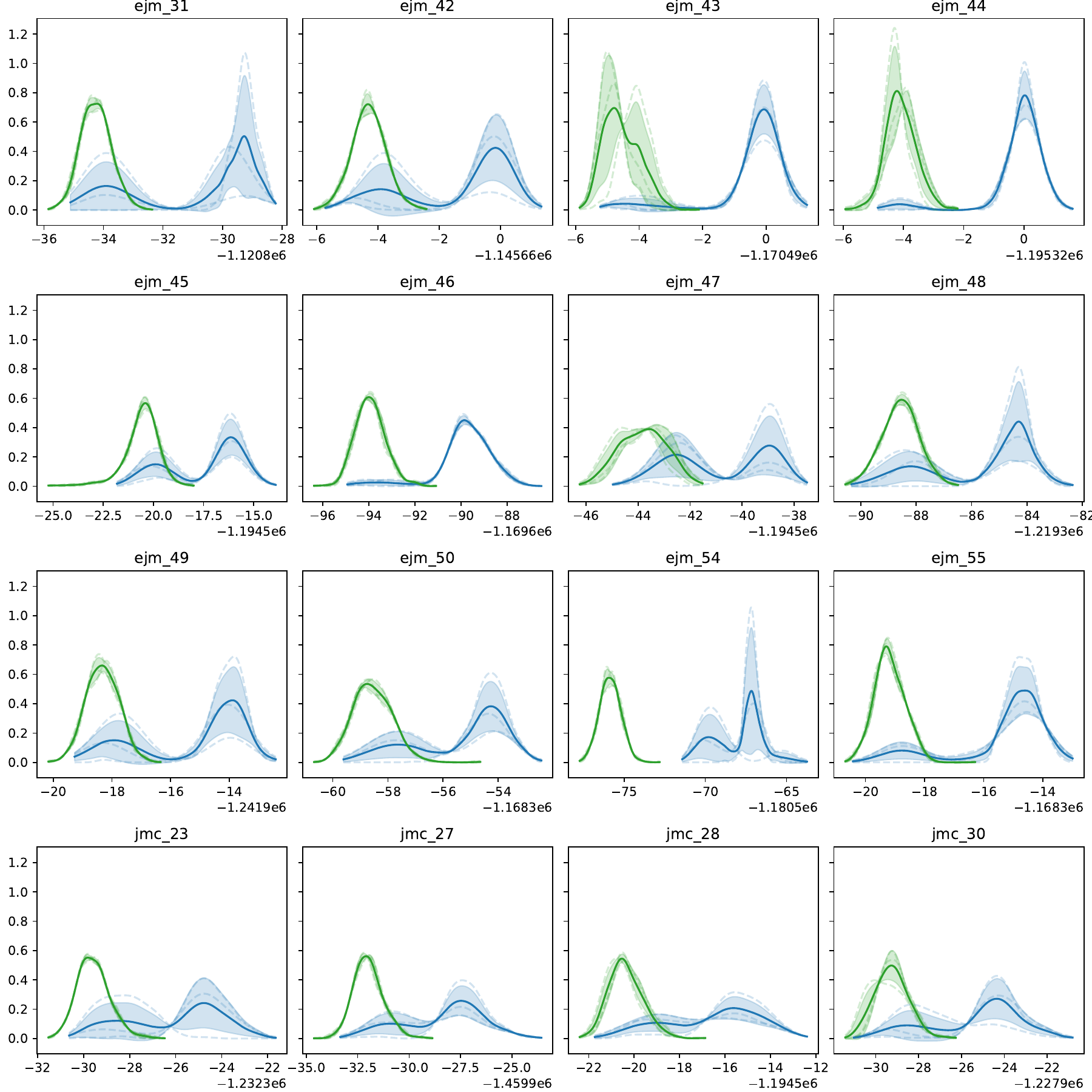}
    \caption{Kernel density plots for the work values of the 300 switches are presented for the \textbf{TYK2-OFF} system in the free leg. Blue indicates the forward direction, switching from MM to ML, while green represents the backward direction (ML to MM). The dashed lines display the density plot for each repeat, whereas the solid blue or green lines represent the average across the three independent runs. The shaded blue and green areas illustrate the standard deviation between the three runs at each point either for the forward (blue) or backwards (green) direction.}
    \label{figS:overlap_tyk2_off}
\end{figure}

\clearpage
\begin{figure}[H]
    \centering
    \includegraphics[width=1\linewidth]{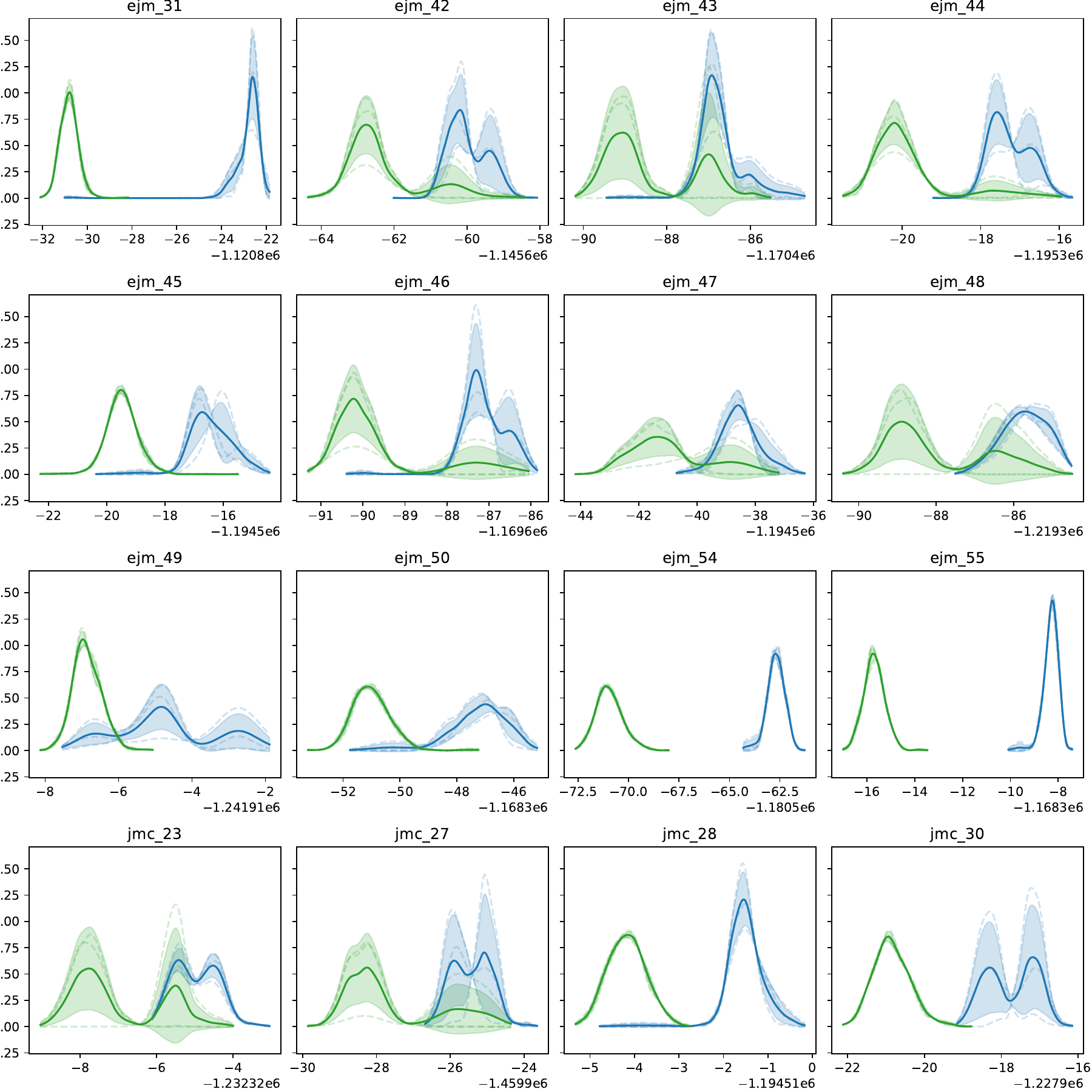}
    \caption{Kernel density plots for the work values of the 300 switches are presented for the \textbf{TYK2-TOR} system in the free leg. Blue indicates the forward direction, switching from MM to ML, while green represents the backward direction (ML to MM). The dashed lines display the density plot for each repeat, whereas the solid blue or green lines represent the average across the three independent runs. The shaded blue and green areas illustrate the standard deviation between the three runs at each point either for the forward (blue) or backwards (green) direction.}
    \label{figS:overlap_tyk2_torp}
\end{figure}
\clearpage

\begin{figure}[H]
    \centering
    \includegraphics[width=1\linewidth]{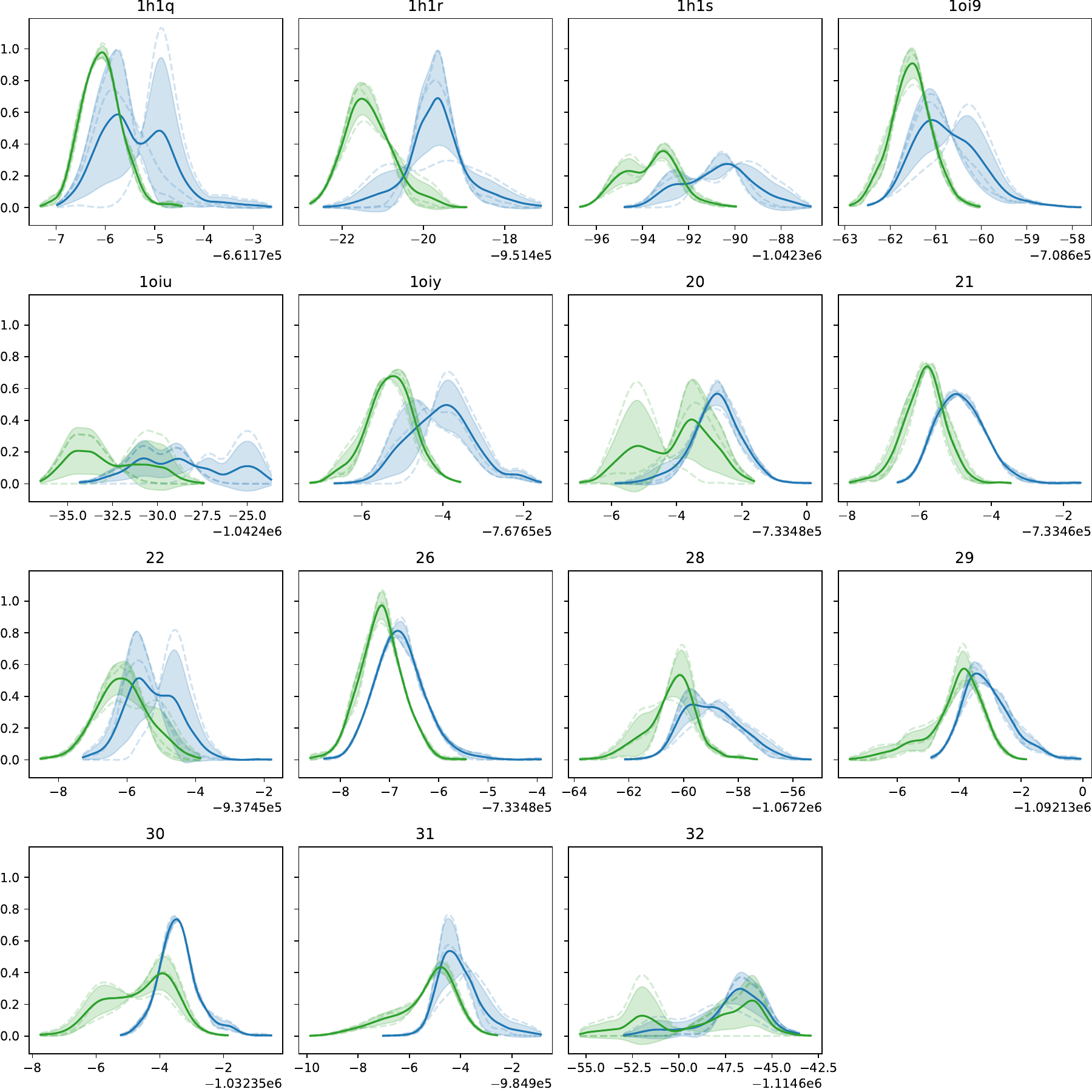}
    \caption{Kernel density plots for the work values of the 300 switches are presented for the \textbf{CDK2-OFF2.2.0} system in the free leg. Blue indicates the forward direction, switching from MM to ML, while green represents the backward direction (ML to MM). The dashed lines display the density plot for each repeat, whereas the solid blue or green lines represent the average across the three independent runs. The shaded blue and green areas illustrate the standard deviation between the three runs at each point either for the forward (blue) or backwards (green) direction.}
    \label{figS:overlap_cdk2_off}
\end{figure}
\clearpage

\begin{figure}[H]
    \centering
    \includegraphics[width=1\linewidth]{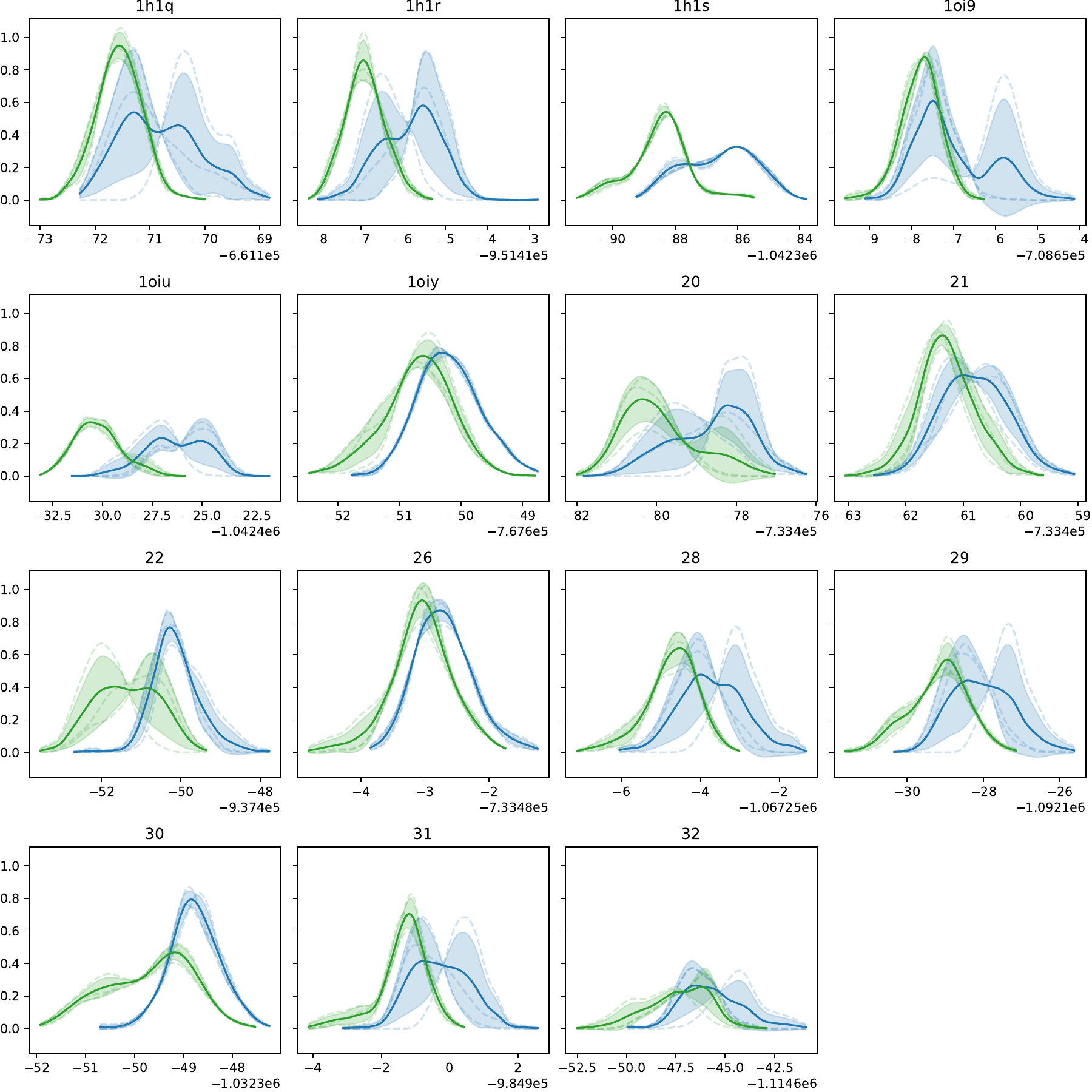}
    \caption{Kernel density plots for the work values of the 300 switches are presented for the \textbf{CDK2-TOR} system in the free leg. Blue indicates the forward direction, switching from MM to ML, while green represents the backward direction (ML to MM). The dashed lines display the density plot for each repeat, whereas the solid blue or green lines represent the average across the three independent runs. The shaded blue and green areas illustrate the standard deviation between the three runs at each point either for the forward (blue) or backwards (green) direction.}
    \label{figS:overlap_cdk2_torp}
\end{figure}
\clearpage

\begin{figure}[H]
    \centering
    \includegraphics[width=0.95\linewidth]{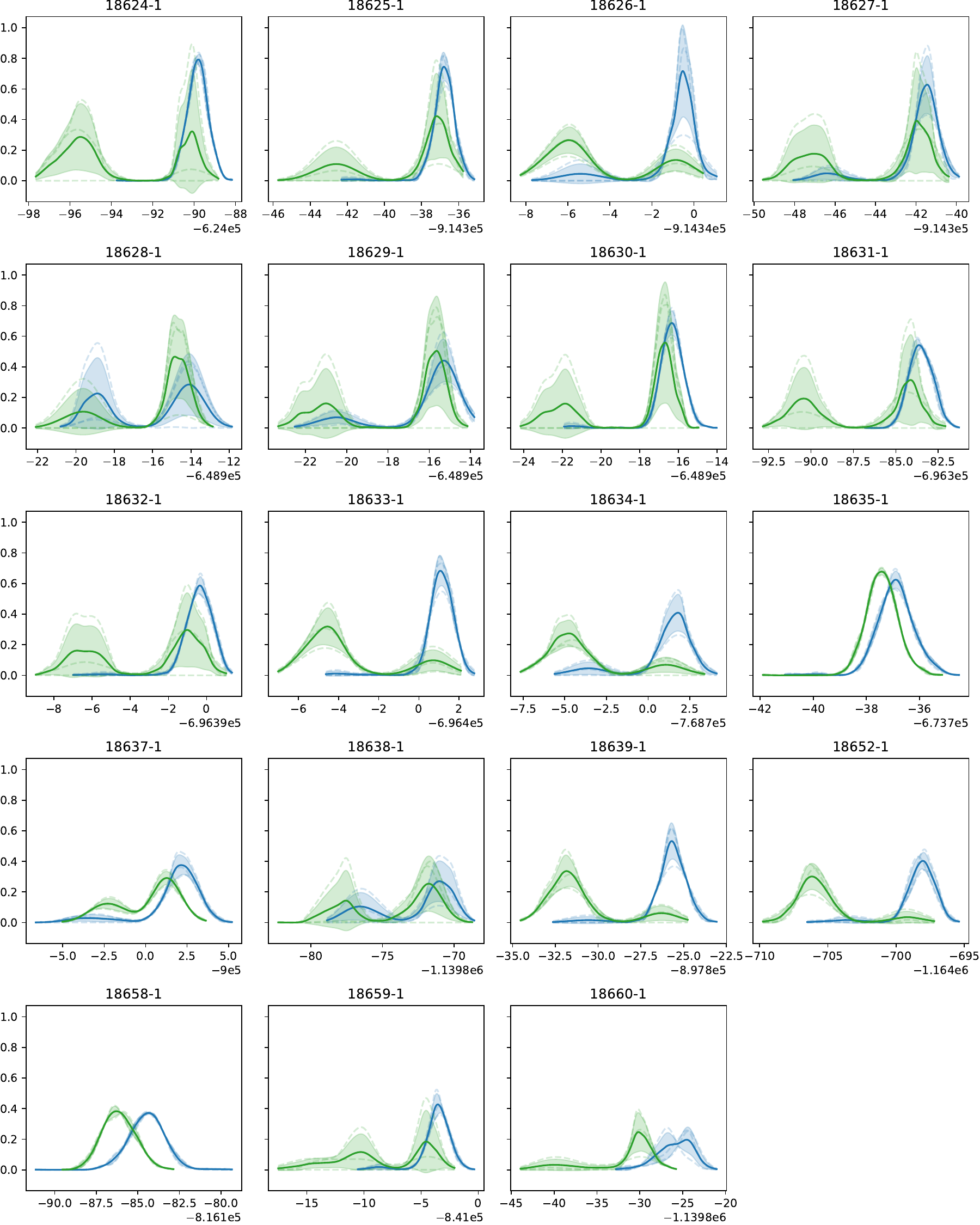}
    \caption{Kernel density plots for the work values of the 300 switches are presented for the \textbf{JNK1-OFF2.2.0} system in the free leg. Blue indicates the forward direction, switching from MM to ML, while green represents the backward direction (ML to MM). The dashed lines display the density plot for each repeat, whereas the solid blue or green lines represent the average across the three independent runs. The shaded blue and green areas illustrate the standard deviation between the three runs at each point either for the forward (blue) or backwards (green) direction.}
    \label{figS:overlap_jnk1_off}
\end{figure}
\clearpage

\begin{figure}[H]
    \centering
    \includegraphics[width=0.95\linewidth]{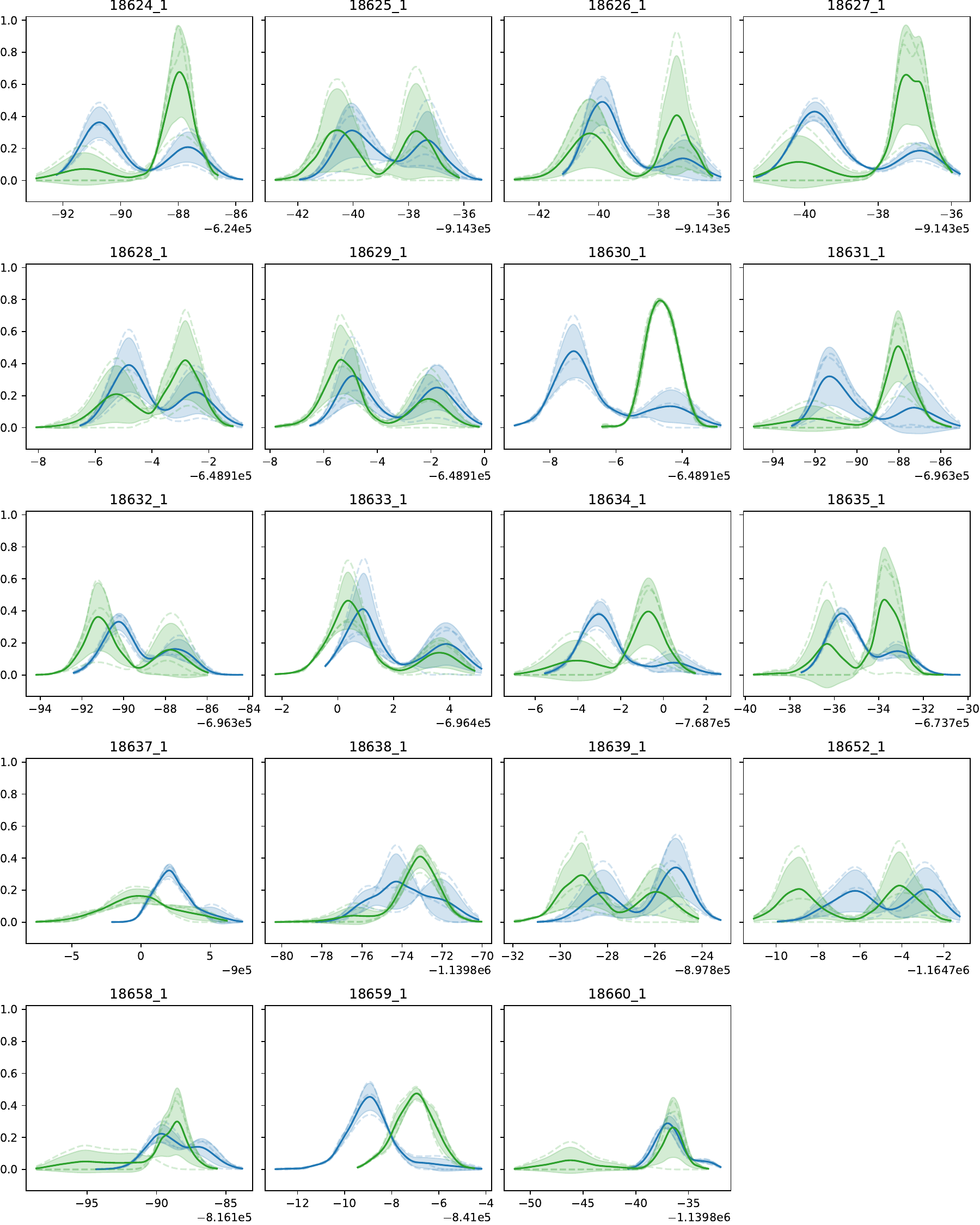}
    \caption{Kernel density plots for the work values of the 300 switches are presented for the \textbf{JNK1-TOR} system in the free leg. Blue indicates the forward direction, switching from MM to ML, while green represents the backward direction (ML to MM). The dashed lines display the density plot for each repeat, whereas the solid blue or green lines represent the average across the three independent runs. The shaded blue and green areas illustrate the standard deviation between the three runs at each point either for the forward (blue) or backwards (green) direction.}
    \label{figS:overlap_jnk1_torp}
\end{figure}
\clearpage

\begin{figure}[H]
    \centering
    \includegraphics[width=0.9\linewidth]{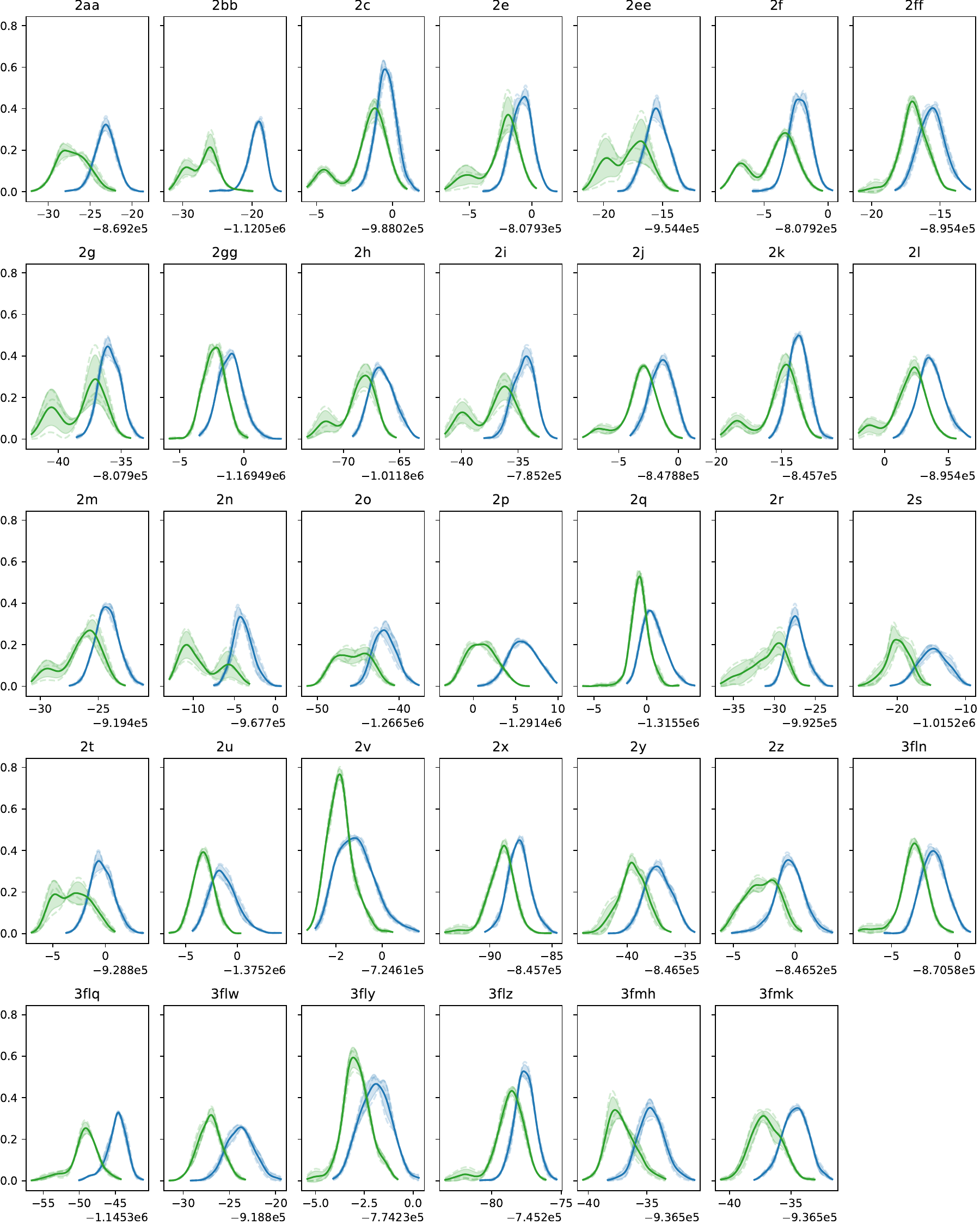}
    \caption{Kernel density plots for the work values of the 300 switches are presented for the \textbf{P38-OFF} system in the free leg. Blue indicates the forward direction, switching from MM to ML, while green represents the backward direction (ML to MM). The dashed lines display the density plot for each repeat, whereas the solid blue or green lines represent the average across the three independent runs. The shaded blue and green areas illustrate the standard deviation between the three runs at each point either for the forward (blue) or backwards (green) direction.}
    \label{figS:overlap_p38_off}
\end{figure}
\clearpage
\begin{figure}[H]
    \centering
    \includegraphics[width=0.95\linewidth]{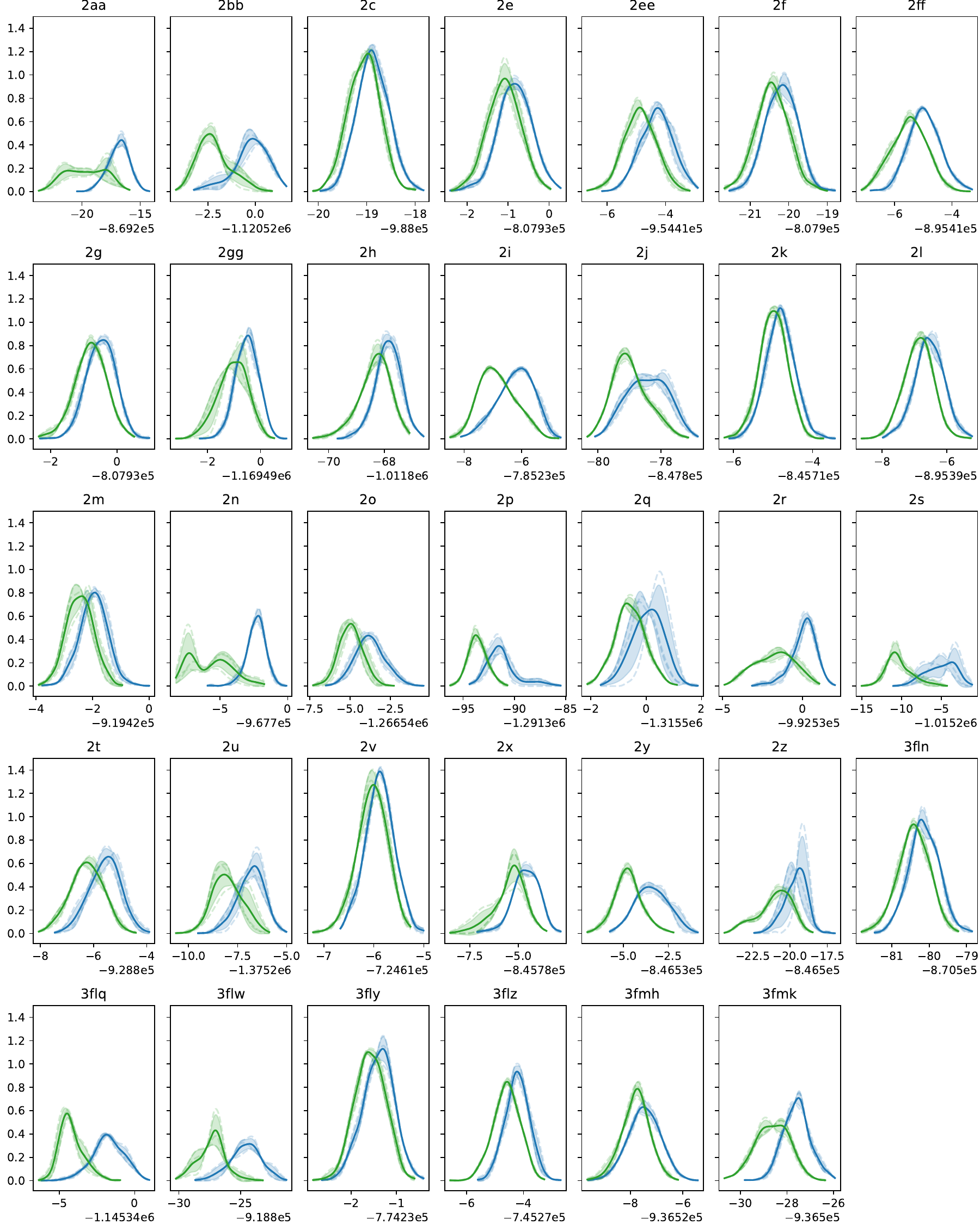}
    \caption{Kernel density plots for the work values of the 300 switches are presented for the \textbf{P38-TOR} system in the free leg. Blue indicates the forward direction, switching from MM to ML, while green represents the backward direction (ML to MM). The dashed lines display the density plot for each repeat, whereas the solid blue or green lines represent the average across the three independent runs. The shaded blue and green areas illustrate the standard deviation between the three runs at each point either for the forward (blue) or backwards (green) direction.}
    \label{figS:overlap_p38_torp}
\end{figure}

\clearpage

\begin{figure}
    \centering
    \includegraphics[width=1\linewidth]{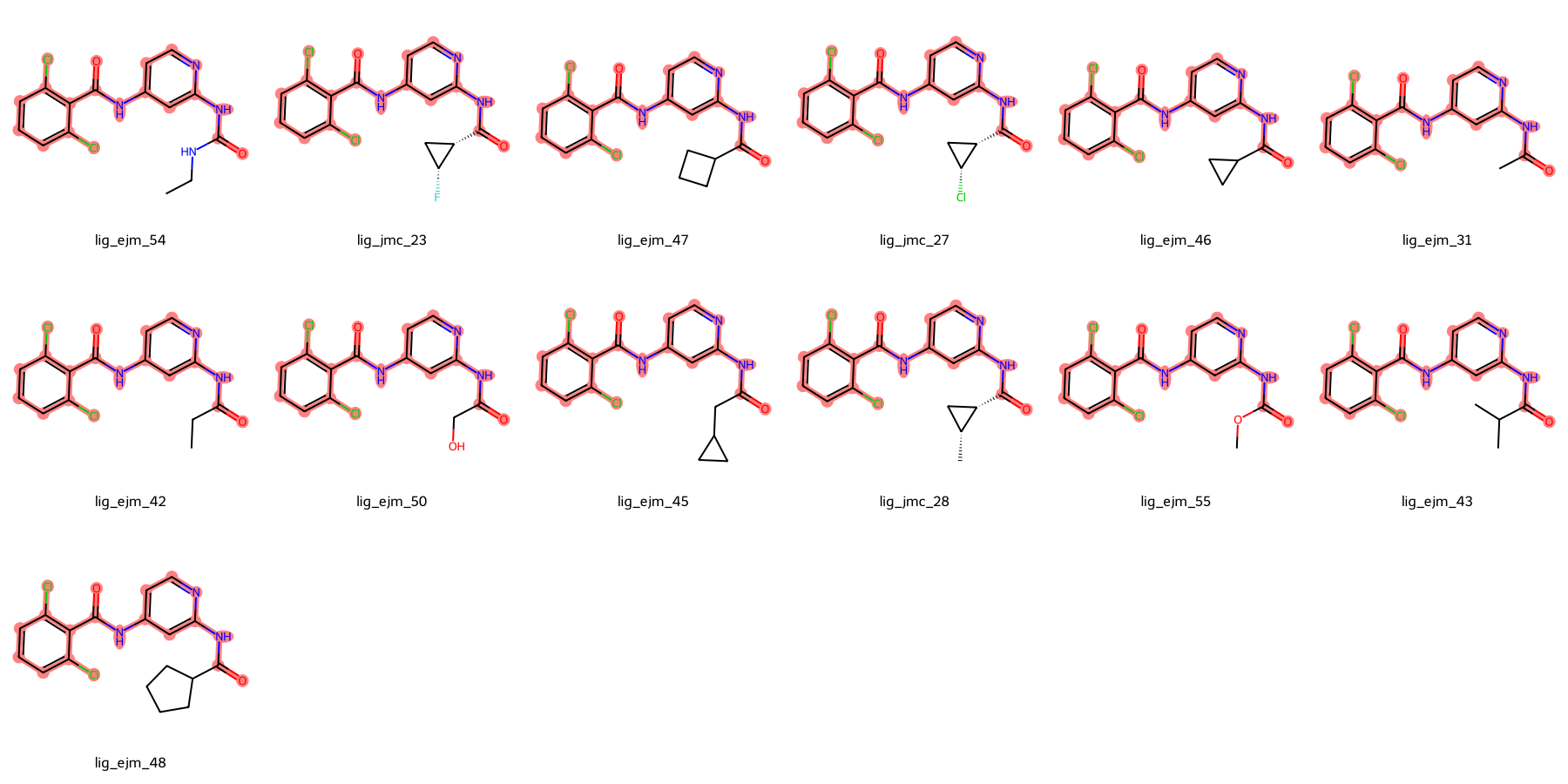}
    \caption{Ligands used in the TYK2 set. The common substructure is highlighted in orange.}
    \label{figS:struc_tyk2}
\end{figure}

\clearpage
\begin{figure}
    \centering
    \includegraphics[width=1\linewidth]{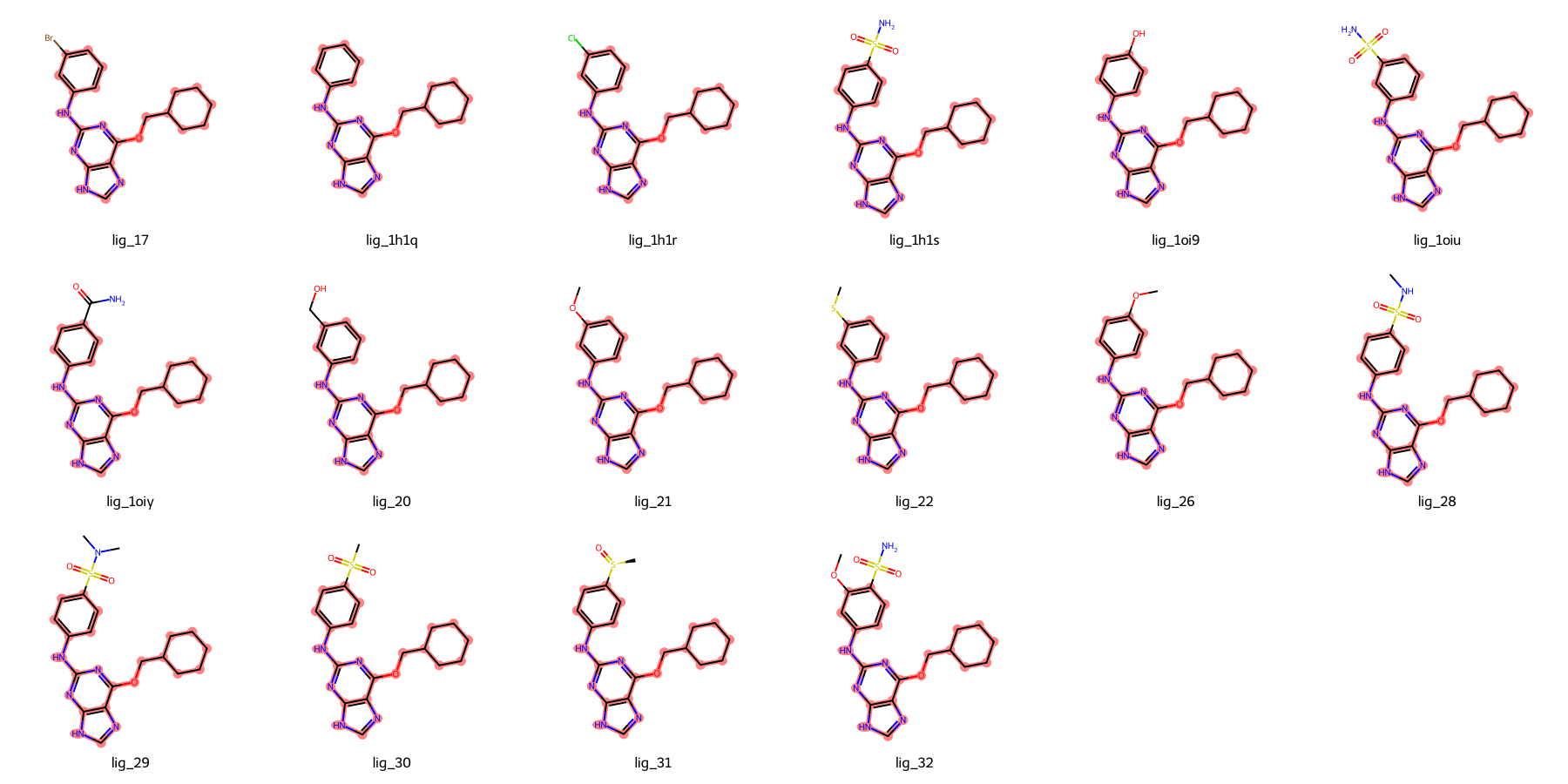}
    \caption{Ligands used in the CDK2 set. The common substructure is highlighted in orange. Ligand 17  had to be excluded because its  Br group is incompatible with the ANI-2x ML potential. }
    \label{figS:struc_cdk2}
\end{figure}
\clearpage
\begin{figure}
    \centering
    \includegraphics[width=1\linewidth]{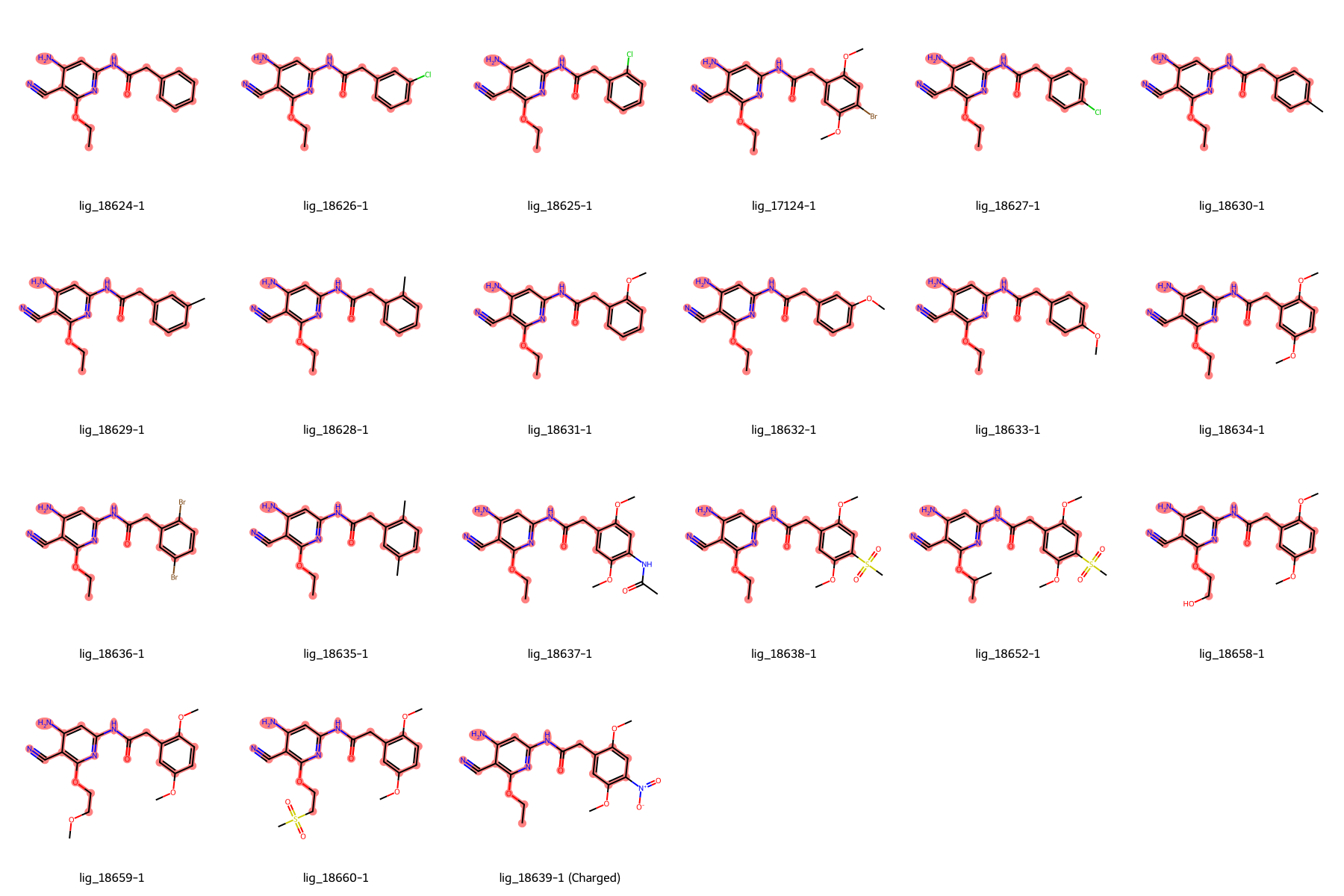}
    \caption{Ligands used in the JNK1 set. The common substructure is highlighted in orange. Ligands 18624 and 18636 had to be excluded because the ANI-2x ML potential does not contain parameters for Br.  Ligand 18639  had to be excluded because its  charged group is incompatible with the ANI-2x ML potential. }
    \label{figS:struc_jnk1}
\end{figure}

\clearpage

\begin{figure}
    \centering
    \includegraphics[width=1\linewidth]{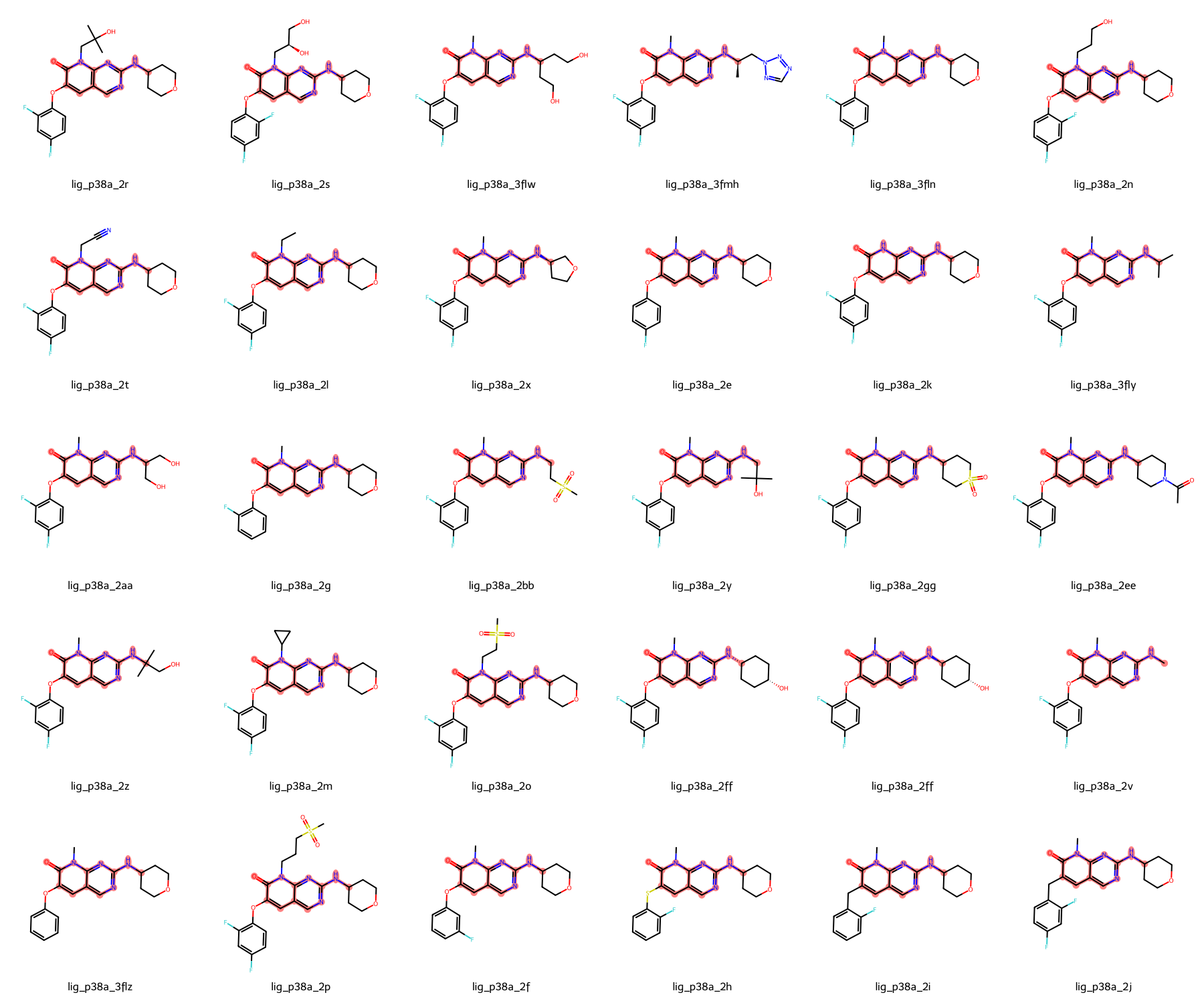}
    \caption{Ligands used in the P38 set. The common substructure is highlighted in orange.}
    \label{figS:struc_p38}
\end{figure}
 
\clearpage
\setstretch{1.0}
